\documentclass[prd,preprint,tightenlines, preprintnumbers,superscriptaddress,amsmath,amssymb,longbibliography]{revtex4-1}

 \usepackage[dvips,final]{graphicx}
  \usepackage{amssymb}
   \usepackage{amsmath}
    \usepackage{amsfonts}
     \usepackage{epsfig}
      \usepackage{bm}
\usepackage{appendix}
\usepackage[section]{placeins}
\usepackage[colorlinks,citecolor=blue,urlcolor=blue,bookmarks=false,hypertexnames=true]{hyperref}

\usepackage{sidecap}
\usepackage{multirow}
\usepackage{booktabs}
\usepackage{array}
\usepackage{tabularx}
\usepackage{xcolor}
\usepackage{pstricks}

\begin{document}

\newcommand{\ds}{\displaystyle}
\newcommand{\mc}{\multicolumn} 
\newcommand{\bce}{\begin{center}}
\newcommand{\ece}{\end{center}}
\newcommand{\beq}{\begin{equation}}
\newcommand{\eeq}{\end{equation}}
\newcommand{\bea}{\begin{eqnarray}}

\newcommand{\eea}{\end{eqnarray}}
\newcommand{\cont}{\nonumber\eea\bea}
\newcommand{\cl}[1]{\begin{center} {#1} \end{center}}
\newcommand{\ea}{\end{array}}

\newcommand{\ab}{{\alpha\beta}}
\newcommand{\cd}{{\gamma\delta}}
\newcommand{\dc}{{\delta\gamma}}
\newcommand{\ac}{{\alpha\gamma}}
\newcommand{\bd}{{\beta\delta}}
\newcommand{\abc}{{\alpha\beta\gamma}}
\newcommand{\eps}{{\epsilon}}
\newcommand{\lam}{{\lambda}}
\newcommand{\mn}{{\mu\nu}}
\newcommand{\mpnp}{{\mu'\nu'}}
\newcommand{\Amuu}{{A_{\mu}}}
\newcommand{\Amuo}{{A^{\mu}}}
\newcommand{\Vmuu}{{V_{\mu}}}
\newcommand{\Vmuo}{{V^{\mu}}}
\newcommand{\Anuu}{{A_{\nu}}}
\newcommand{\Anuo}{{A^{\nu}}}
\newcommand{\Vnuu}{{V_{\nu}}}
\newcommand{\Vnuo}{{V^{\nu}}}
\newcommand{\Fmnu}{{F_{\mu\nu}}}
\newcommand{\Fmno}{{F^{\mu\nu}}}

\newcommand{\shat}{{\hat{s}}}
\newcommand{\that}{{\hat{t}}}
\newcommand{\uhat}{{\hat{u}}}

\newcommand{\abcd}{{\alpha\beta\gamma\delta}}


\newcommand{\bsigma}{\mbox{\boldmath $\sigma$}}
\newcommand{\beps}{\mbox{\boldmath $\varepsilon$}}
\newcommand{\btau}{\mbox{\boldmath $\tau$}}
\newcommand{\brho}{\mbox{\boldmath $\rho$}}
\newcommand{\bpipi}{\mbox{\boldmath $\pi\pi$}} 
\newcommand{\bss}{\bsigma\!\cdot\!\bsigma}
\newcommand{\btt}{\btau\!\cdot\!\btau}
\newcommand{\bnabla}{\mbox{\boldmath $\nabla$}}
\newcommand{\bphi}{\mbox{\boldmath $\tau$}}
\newcommand{\bvarphi}{\mbox{\boldmath $\rho$}}
\newcommand{\bE}{\mbox{\boldmath $E$}}
\newcommand{\bDelta}{\mbox{\boldmath $\Delta$}}
\newcommand{\bGamma}{\mbox{\boldmath $\Gamma$}}
\newcommand{\bpsi}{\mbox{\boldmath $\psi$}}
\newcommand{\bPsi}{\mbox{\boldmath $\Psi$}}
\newcommand{\bPhi}{\mbox{\boldmath $\Phi$}}
\newcommand{\bnab}{\mbox{\boldmath $\nabla$}}
\newcommand{\bpi}{\mbox{\boldmath $\pi$}}
\newcommand{\btheta}{\mbox{\boldmath $\theta$}}
\newcommand{\bkappa}{\mbox{\boldmath $\kappa$}}
\newcommand{\bgamma}{\mbox{\boldmath $\gamma$}}

\newcommand{\bp}{\mbox{\boldmath $p$}}
\newcommand{\ba}{\mbox{\boldmath $a$}}
\newcommand{\bq}{\mbox{\boldmath $q$}}
\newcommand{\br}{\mbox{\boldmath $r$}}
\newcommand{\bs}{\mbox{\boldmath $s$}}
\newcommand{\bk}{\mbox{\boldmath $k$}}
\newcommand{\bl}{\mbox{\boldmath $l$}}
\newcommand{\bb}{\mbox{\boldmath $b$}}
\newcommand{\be}{\mbox{\boldmath $e$}}
\newcommand{\bP}{\mbox{\boldmath $P$}}
\newcommand{\bV}{\mbox{\boldmath $V$}}
\newcommand{\bI}{\mbox{\boldmath $I$}}
\newcommand{\bJ}{\mbox{\boldmath $J$}}

\newcommand{\bT}{{\bf T}}
\newcommand{\fph}{${\cal F}$}
\newcommand{\aph}{${\cal A}$}
\newcommand{\dph}{${\cal D}$}
\newcommand{\fpi}{f_\pi}
\newcommand{\mpi}{m_\pi}
\newcommand{\Tr}{{\mbox{\rm Tr}}}
\def\Qb{\overline{Q}}
\newcommand{\delu}{\partial_{\mu}}
\newcommand{\delo}{\partial^{\mu}}
\newcommand{\up}{\!\uparrow}
\newcommand{\upup}{\uparrow\uparrow}
\newcommand{\updo}{\uparrow\downarrow}
\newcommand{\uu}{$\uparrow\uparrow$}
\newcommand{\ud}{$\uparrow\downarrow$}
\newcommand{\auu}{$a^{\uparrow\uparrow}$}
\newcommand{\aud}{$a^{\uparrow\downarrow}$}
\newcommand{\pu}{p\!\uparrow}
\newcommand{\qp}{quasiparticle}
\newcommand{\sa}{scattering amplitude}
\newcommand{\ph}{particle-hole}
\newcommand{\qcd}{{\it QCD}}
\newcommand{\integ}{\int\!d}
\newcommand{\ie}{{\sl i.e.~}}
\newcommand{\etal}{{\sl et al.~}}
\newcommand{\etc}{{\sl etc.~}}
\newcommand{\rhs}{{\sl rhs~}}
\newcommand{\lhs}{{\sl lhs~}}
\newcommand{\eg}{{\sl e.g.~}}
\newcommand{\ef}{\epsilon_F}
\newcommand{\sigt}{d^2\sigma/d\Omega dE}
\newcommand{\sige}{{d^2\sigma\over d\Omega dE}}
\newcommand{\rpaeq}{\beq
\left ( \begin{array}{cc}
A&B\\
-B^*&-A^*\end{array}\right )
\left ( \begin{array}{c}
X^{(\kappa})\\Y^{(\kappa)}\end{array}\right )=E_\kappa
\left ( \begin{array}{c}
X^{(\kappa})\\Y^{(\kappa)}\end{array}\right )
\eeq}

\newcommand{\ket}[1]{{#1} \rangle}
\newcommand{\bra}[1]{\langle {#1} }

\newcommand{\Bigket}[1]{{#1} \Big\rangle}
\newcommand{\Bigbra}[1]{\Big\langle {#1} }

\newcommand{\ave}[1]{\langle {#1} \rangle}
\newcommand{\Bigave}[1]{\left\langle {#1} \right\rangle}
\newcommand{\half}{{\frac{1}{2}}}

\newcommand{\singlespace}{
    \renewcommand{\baselinestretch}{1}\large\normalsize}
\newcommand{\doublespace}{
    \renewcommand{\baselinestretch}{1.6}\large\normalsize}
\newcommand{\bftau}{\mbox{\boldmath $\tau$}}
\newcommand{\bfalpha}{\mbox{\boldmath $\alpha$}}
\newcommand{\bfgamma}{\mbox{\boldmath $\gamma$}}
\newcommand{\bfxi}{\mbox{\boldmath $\xi$}}
\newcommand{\bfbeta}{\mbox{\boldmath $\beta$}}
\newcommand{\bfeta}{\mbox{\boldmath $\eta$}}
\newcommand{\bfpi}{\mbox{\boldmath $\pi$}}
\newcommand{\bfphi}{\mbox{\boldmath $\phi$}}
\newcommand{\bfR}{\mbox{\boldmath ${\cal R}$}}
\newcommand{\bfL}{\mbox{\boldmath ${\cal L}$}}
\newcommand{\bfM}{\mbox{\boldmath ${\cal M}$}}
\def\dblint{\mathop{\rlap{\hbox{$\displaystyle\!\int\!\!\!\!\!\int$}}
    \hbox{$\bigcirc$}}}
\def\ut#1{$\underline{\smash{\vphantom{y}\hbox{#1}}}$}

\def\UNITY{{\bf 1\! |}}
\def\Pom{{\bf I\!P}}
\def\lsim{\mathrel{\rlap{\lower4pt\hbox{\hskip1pt$\sim$}}
    \raise1pt\hbox{$<$}}}         
\def\gsim{\mathrel{\rlap{\lower4pt\hbox{\hskip1pt$\sim$}}
    \raise1pt\hbox{$>$}}}         

\newcommand\scalemath[2]{\scalebox{#1}{\mbox{\ensuremath{\displaystyle #2}}}}

\newcommand{\RP}[1]{{\blue RP: #1}}
\newcommand{\WS}[1]{{\red WS: #1}}

\title{Two photon decay width of the fully charmed tetraquarks: revisiting prospects for ultraperipheral collisions
}

\author{Longjie Chen}
\email{longjie.chen@ifj.edu.pl}
\affiliation{Institute of Nuclear Physics, Polish Academy of Sciences, 
ul. Radzikowskiego 152, PL-31-342 Krak{\'o}w, Poland}

\author{Wolfgang Sch\"afer}%
\email{wolfgang.schafer@ifj.edu.pl}
\affiliation{Institute of Nuclear
Physics, Polish Academy of Sciences, ul. Radzikowskiego 152, PL-31-342 
Krak{\'o}w, Poland}

\author{Antoni Szczurek}%
\email{antoni.szczurek@ifj.edu.pl}
\affiliation{Institute of Nuclear
Physics, Polish Academy of Sciences, ul. Radzikowskiego 152, PL-31-342 
Krak{\'o}w, Poland}
\affiliation{Institute of Physics, University of Rzeszów, ul. Pigonia 1, PL-35-959 Rzeszów, Poland}
\begin{abstract}
We discuss the role of fully heavy tetraquarks in ultraperipheral collisions $AA\to AA\, J/\psi J/\psi$ and $AA \to AA\, \gamma \gamma$. Two-photon couplings to scalar and tensor tetraquarks are considered. We use relatively recent results of four-body calculation of fully heavy tetraquark wave function within the extended relativized quark model. The corresponding radiative decay widths for different tetraquark states are evaluated using NRQCD factorization, with LDMEs extracted from the four-body wave functions at the origin. The results are collected in tables. The cross sections for production of pairs of $J/\psi$ mesons and diphotons in UPC of $^{208}Pb + {^{208}Pb}$ collisions are presented. While the couplings for $T_{4c}\left(0^{++}, 2^{++}\right) \to \gamma \gamma$ are calculated based on the model wave functions, simplified couplings are used for $T_{4c}\left(0^{++}, 2^{++}\right) \to J/\psi J/\psi$. When calculating the energy dependence of the resonant $\gamma \gamma \to J/\psi J/\psi$ and $\gamma \gamma \to \gamma \gamma$ cross section, we employ the total decay widths measured at $\Gamma_{\rm{tot}}=0.446\,\rm{GeV}$ for $X(6600)$ while $\Gamma_{\rm{tot}}=0.135\,\rm{GeV}$ for $X(6900)$ in latest CMS data. The resonant terms are compared with continuum contributions for both considered channels. While for the $J/\psi J/\psi$ channel the resonant contributions are larger than continuum ones, for the $\gamma \gamma$ channels the situation is reversed. The latter result is in clear disagreement with the result obtained from the n\"aive use of vector dominance picture.
\end{abstract}

\maketitle

\section{Introduction}

The observation of resonant structure in the $J/
\psi$--pair mass spectrum by the LHCb collaboration \cite{LHCb:2020bwg}, later confirmed by ATLAS \cite{ATLAS:2023bft,ATLAS:2025nsd} and CMS \cite{CMS:2023owd,CMS:2025fpt,CMS:2026tiu} in the $J/\psi J/\psi$ and $J/\psi \psi'$ channels has sparked interest in fully heavy--quark exotic states. The discovery of this structure, $X( 6900)$, provides the first definitive evidence for the existence of $cc\bar{c}\bar{c}$ tetraquarks. 

Its fundamental nature remains an open question, with the leading hypothesis being a compact diquark--antidiquark bound state, where the constituents are organised into coloured clusters. 
Recent measurements by CMS \cite{CMS:2025fpt} prefer the $J^{PC} = 2^{++}$ assignment for this state. For a summary of states observed by the LHC experiments in the $J/\psi J/\psi$ channel and their extracted parameters, see Table~\ref{tab:T4c_states}.
\begin{table}
\caption{
Parameters (mass $M$ and decay width $\Gamma$) of $T_{4c}$ states as extracted by LHC experiments from $J/\psi J/\psi$-invariant mass spectra. All masses and widths are given in MeV.
}
\centering
\begin{ruledtabular}
\begin{tabular}{ccccc}
State &    & LHCb \cite{LHCb:2020bwg}& ATLAS \cite{ATLAS:2023bft} & CMS \cite{CMS:2026tiu} \\
\hline
X(6400) & $M$ &  &  $6410 \pm 80 ^{+80}_{-30}$ &  \\
& $\Gamma$ & & $590 \pm 350^{+120}_{-200}$ &   \\
\hline
X(6600) & $M$ &  &  $6630 \pm 50^{+80}_{-10}$    &  $6593^{+15}_{-14} \pm 25$ \\
& $\Gamma$ &  &  $350 \pm 110^{+100}_{-40}$   & $446^{+66}_{-54} \pm 87$\\
\hline
X(6900) & $M$ & $6905 \pm 11 \pm 7$ & $6860 \pm 30^{+10}_{-20}$  & $6847 \pm 10 \pm 15$ \\ 
& $\Gamma$ & $80 \pm 19 \pm 33$ &
$110 \pm 50^{+20}_{-10}$ & $135^{+16}_{-14} \pm 14$ \\
\hline
X(7100) & $M$ &  &  & $7173^{+9}_{-10} \pm 13$\\
& $\Gamma$ &  &  & $73^{+18}_{-15} \pm 10$ \\
\end{tabular}
\end{ruledtabular}
\label{tab:T4c_states}
\end{table}

While hadroproduction in $pp$ collisions has been the primary mode of discovery, the interpretation of these results is complicated by the complex hadronic environment. Theoretical studies of production mechanisms in this mode are extensive \cite{Berezhnoy:2011xy,Maciula:2020wri,Feng:2020riv,Zhu:2020xni,Feng:2023agq,
Celiberto:2024mab,
Belov:2024qyi,Celiberto:2025ziy,Wang:2025hex,Celiberto:2025vra}, but potential contributions from processes such as double parton scattering can obscure the intrinsic properties of the tetraquark itself. Therefore, a cleaner experimental channel is highly desirable to probe its internal structure directly.

The exclusive process $\gamma\gamma\to X( 6900 )$ serves as such a channel. It is accessible via photon-photon scattering in ultraperipheral heavy-ion collisions (UPCs), where the colliding ions act as intense sources of quasi-real photons. The production of conventional quarkonium pairs, such as $J/\psi$ pairs, in UPCs has been previously considered \cite{Baranov:2012vu,Yang:2025vcs}, and the potential for discovering and studying exotic tetraquark states in this clean environment has been recently highlighted \cite{Goncalves:2021ytq,Esposito:2021ptx,Biloshytskyi:2022dmo,dEnterria:2025ecx}. The key advantage is that the cross section for this exclusive process is directly proportional to the two-photon decay width, $\Gamma_{\gamma\gamma}$, a quantity that is highly sensitive to the internal configuration of the state. Considerations of tetraquark radiative and other decay widths can be found, for example, in \cite{Sang:2023ncm,Kalamidas:2025gen,Becchi:2020uvq,Zhang:2023ffe}. A measurement of the decay width can thus act as a powerful discriminant between different configurations of internal structure.

Here we present estimates of the two--photon decay widths for both scalar ($0^{++}$) and tensor ($2^{++}$) fully heavy tetraquarks at the lowest order in the NRQCD approach. By calculating these widths within a compact diquark-antidiquark framework, we provide a clear benchmark for future experimental measurements at the LHC, which will be crucial in elucidating the true nature of the $X( 6900 )$. The work is organized as follows: 
In Section~\ref{sec:gammagamma} we present our calculation of the two--photon couplings for $0^{++}$ and $2^{++}$ fully $QQ\bar Q \bar Q$ tetraquark $S$--wave states, adopting NRQCD factorization. In Section \ref{section:LDMEs} we present our evaluation of the relevant long--distance matrix elements (LDMEs) based on quark--model wave functions, and we present our results for the two--photon decay width of the fully charmed tetraquark states.
In Section~\ref{sec:UPC} we turn to the phenomenology of tetraquark production in UPCs via photon--photon fusion, and discuss the prospects for measurement in $J/\psi J/\psi$ as well as $\gamma \gamma$ final states. We present our conclusions in Section~\ref{sec:conclusions}.

\section{Two--photon couplings}
\label{sec:gammagamma}
\begin{figure}
\includegraphics[width=.8\textwidth]{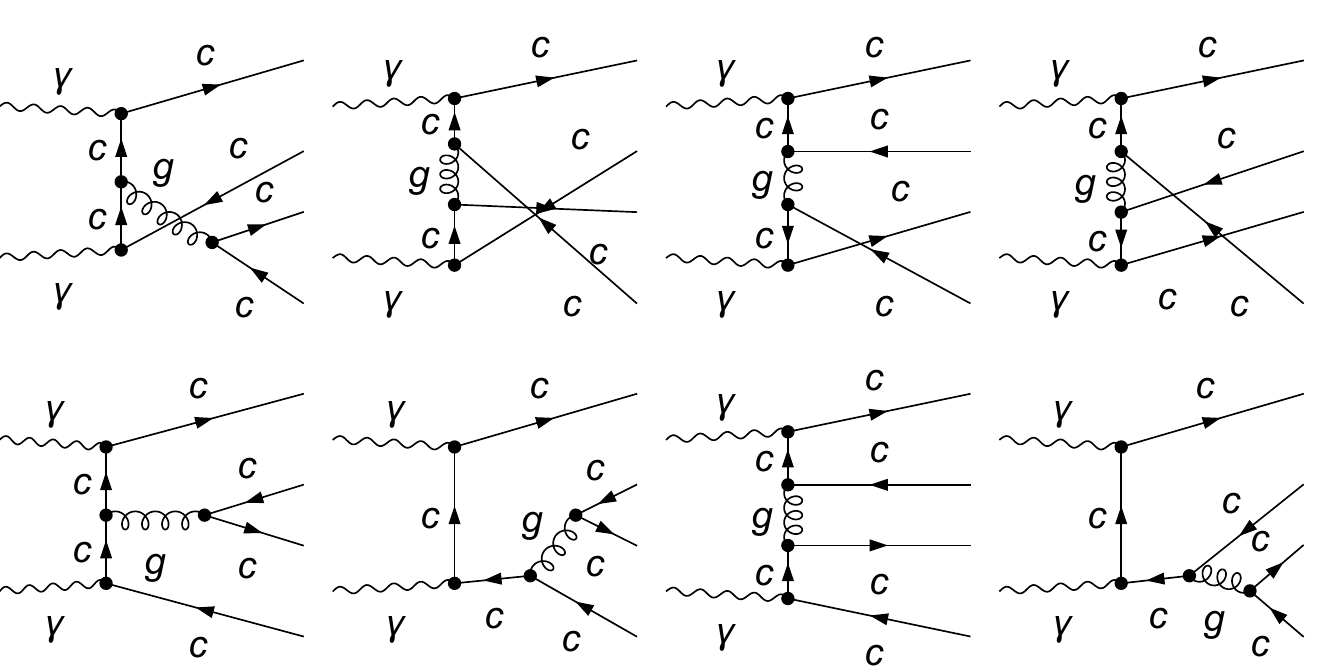}
\caption{Some of the 40 Feynman diagrams for the $\gamma \gamma \to c c \bar c \bar c$ amplitude.}
\label{fig:diagram1}
\end{figure}
We start by reminding the reader of the two--photon couplings for the scalar and tensor particles \cite{Poppe:1986dq,Budnev:1975poe}. 
\begin{eqnarray}
    {\cal M}^{(0)}_{\mu \nu} &=& G_{\mu \nu} \, e^2  F_{\rm TT}, \, \nonumber  \\
    {\cal M}^{(2)}_{\mu \nu}(J_z) &=& G_{\mu \nu} (q_1 - q_2)_\alpha (q_1 - q_2)_\beta E^{* \alpha \beta}(J_z) \,e^2 F_{\rm TT,0} \nonumber \\
    &+& \half \Big(G_{\mu \alpha} G_{\nu \beta} + G_{\mu \beta} G_{\nu \alpha} - G_{\mu \nu} G_{\alpha \beta} \Big) E^{* \alpha \beta}(J_z) \,  e^2 F_{\rm TT,2} \, , 
\end{eqnarray}
where $e^2 = 4 \pi \alpha_{\rm em}$.
Here we have defined
\begin{eqnarray}
    G_{\mu \nu} = - g_{\mu \nu} + \frac{q_{1\mu}q_{2\nu} + q_{2\mu} q_{1\nu}}{q_1 \cdot q_2} \, , 
\end{eqnarray}
and our amplitudes are manifestly gauge invariant $q_1^{
\mu} {\cal M}_{\mu \nu} = q_2^\nu {\cal M}_{\mu \nu} = 0$.
In this work we will only discuss the case of on--shell photons, $q_1^2 = q_2^2 = 0$.

The two--photon decay widths of $0^{++}$ and $2^{++}$ states are given in terms of the  couplings $F_i$ as:
\begin{eqnarray}
    \Gamma(0^{++} \to \gamma \gamma) &=&  \frac{\pi \alpha_{\rm em}^2}{M} |F_{\rm TT}|^2, \nonumber \\
     \Gamma(2^{++} \to \gamma \gamma) &=& \frac{\pi \alpha_{\rm em}^2}{5 M} \Big( |F_{\rm TT,2}|^2 + \frac{2}{3} M^4 |F_{\rm TT,0}|^2 
    \Big) .
\end{eqnarray}
Here $M$ denotes the mass of the respective particle.

For our estimate of the two-photon couplings, we employ NRQCD factorization~\cite{Bodwin:1994jh}. The amplitude for $\gamma\gamma \to T_{4c}$ is assumed to factorize into perturbative short-distance coefficients (SDCs) describing the production of a $c\bar{c}c\bar{c}$ system at the hard scale $m_c$, and LDMEs encoding the non-perturbative transition to the bound state. Representative Feynman diagrams contributing to the short-distance process at tree level, generated with \textsc{FeynCalc}~\cite{Shtabovenko:2023idz}, are shown in Fig.~\ref{fig:diagram1}.

We start from the assumption, that the fully heavy tetraquark state under consideration can be described as a compact diquark--antidiquark bound state. In this picture, the tetraquark is composed of a diquark$(cc)$ and an anti--diquark$( \bar{c}\bar{c})$, each with a well-defined color and spin structure. We assume that each of the diquarks is in a relative $S$-wave state. 

Furthermore, we assume that the relative orbital angular momentum between the diquark and the anti-diquark is also zero, meaning the overall tetraquark is an $S$-wave state. This configuration maximizes the overlap of the constituent quark wave functions and is expected to be the dominant component. In the language of NRQCD, this $S$-wave nature means that the leading-order LDMEs are proportional to the square of the tetraquark's wave function at the origin. The diquarks can be formed in different color representations, which are then combined to form a color-singlet tetraquark. Notice, that the coupling of diquarks into good quantum numbers does not imply a quasi--two body structure of the bound states. Rather it puts into evidence the role of the Pauli principle for the relevant configurations.

We thus construct the NRQCD operators by first coupling $QQ$ and $\bar Q \bar Q$ diquarks and antidiquarks as
\begin{table}[]
    \centering
    \begin{tabular}{c|c|c}
       Operator  & Color & Spin$^{
       \rm Parity}$       \\
\hline
      ${\cal D}_{kl}$   & ${\bf 6}_S$ & $0^+_A$ \\
      ${\overline{
      \cal D}}_{kl}$ & ${\overline{
      \bf 6}}_S$ & $0^{+}_A$ \\
      ${\cal D}^i_{kl}$   & $\overline {\bf 3}_A$ & $1^+_S$   \\
      $\overline{\cal D}^i_{kl}$   & $ {\bf 3}_A$ & $1^+_S$    
      \end{tabular}
    \caption{Color representation and spin/parity of diquark operators, also shown are the symmetry properties under particle exchange.}
    \label{tab:diquarks}
\end{table}
\begin{eqnarray}
    {\cal D}_{kl} &=& {\cal S}_{kl,rs} \, \psi^T_r \, i \sigma^2 \, \psi_s, \quad   \overline {\cal D}_{kl} =  {\cal S}_{kl,rs} \, \chi^{\dagger}_r \, i \sigma^2 \, \chi^{*}_s \nonumber \\
    {\cal D}^j_{kl} &=& {\cal A}_{kl,rs} \psi^T_r \, i \sigma^2 \sigma^j \, \psi_s, \quad  \overline {\cal D}^j_{kl} =  {\cal A}_{kl,rs} \, \chi^{\dagger}_r \,\sigma^j i \sigma^2 \, \chi^{*}_s, 
\end{eqnarray}
with
\begin{eqnarray}
  {\cal S}_{kl,rs}  &=& \sqrt{\frac{2}{N_c (N_c + 1)}} \, \half \Big( \delta_{kr} \delta_{ls} + \delta_{ks}\delta_{lr} \Big) , \nonumber \\ 
    {\cal A}_{kl,rs}  &=& \sqrt{\frac{2}{N_c (N_c - 1)}} \, \half \Big( \delta_{kr} \delta_{ls} - \delta_{ks}\delta_{lr} \Big) \, .
\end{eqnarray}
The symmetry properties of these diquark operators are summarized in Table~\ref{tab:diquarks}. As the total wavefunction of the two (anti--)fermion system must be antisymmetric, we see that for the $S$--wave, the color (anti-)sextet must be accompanied by the symmetric spin-singlet state, while the color (anti--) triplet is in a spin--triplet state. 
From these diquark operators, we go on to construct the color singlet tetraquark operators
\begin{eqnarray}
    {\cal O}^{(0)}_{\bf 6 \otimes \bar 6} = {
    \cal D}_{kl} \overline {\cal D}_{kl}, \quad  {\cal O}^{(0)}_{\bf \bar 3 \otimes  3} =  \frac{1}{\sqrt{3}} \delta_{ij}  {
    \cal D}^i_{kl} \overline {\cal D}^j_{kl}, \quad 
    {\cal O}^{(2)}_{\bf \bar 3 \otimes  3} =  E^*_{ij}(J_z)  {
    \cal D}^i_{kl} \overline {\cal D}^j_{kl} \, .
\end{eqnarray}
The LDMEs $\bra{0}| {\cal O}^{(J)}_{\bf R \otimes \bar R}|\ket{T^{J}_{4Q}}$ will later be related to the four--body wavefunctions at the origin for the relevant spin--color configuration. 
For the perturbative LDMEs, we have that 
\begin{eqnarray}
    \bra{0} | {\cal O}^{(0)}_{\bf 6 \otimes \bar 6} | \ket{[Q Q]^0_{\bf 6} [\bar Q \bar Q]^0_{\bf \bar 6}} =    \bra{0} | {\cal O}^{(0)}_{\bf \bar 3 \otimes 3} | \ket{[Q Q]^1_{\bf \bar 3} [\bar Q \bar Q]^1_{\bf 3}}=
    \bra{0} | {\cal O}^{(2)}_{\bf \bar 3 \otimes 3} | \ket{[Q Q]^1_{\bf \bar 3} [\bar Q \bar Q]^1_{\bf 3}} = 4 (2m_Q)^2 \, . 
\end{eqnarray}
The NRQCD factorization formula for the two--photon couplings for the $0^{++}$ state reads
\begin{eqnarray}
   F_{\rm TT} = \tilde F^{{\bf 6} \overline{\bf 6}}_{\rm TT} \cdot \frac{\sqrt{2M} \bra{0}| O^{(0)}_{\bf 6 \otimes \bar 6} |\ket{T_{4Q}}}{4(2 m_Q)^2}  + \tilde F^{\overline{\bf 3} {\bf 3}}_{
   \rm TT} 
   \cdot \frac{\sqrt{2M} \bra{0}|{\cal O}^{(0)}_{ \bf \bar 3 \otimes \bf 3}
    |\ket{T_{4Q}}}{4(2m_Q)^2}, 
\end{eqnarray}
while for the $2^{++}$ state, we have
\begin{eqnarray}
F_{\rm TT,0} = \tilde{F}_{\rm TT,0} \, \frac{\sqrt{2M} \bra{0}|{\cal O}^{(2)}_{ \bf \bar 3 \otimes \bf 3}
    |\ket{T_{4Q}}}{4(2m_Q)^2}, \quad
F_{\rm TT,2} = \tilde{F}_{\rm TT,2} \, \frac{\sqrt{2M} \bra{0}|{\cal O}^{(2)}_{ \bf \bar 3 \otimes \bf 3}
    |\ket{T_{4Q}}}{4(2m_Q)^2}. 
\end{eqnarray}

The SDCs are perturbatively calculable, which are determined by performing a matching calculation at leading order (LO) between the full QCD theory and the NRQCD effective theory. This procedure involves computing the tree-level Feynman diagrams for the partonic subprocess shown as Fig.\ref{fig:diagram1}.

For the tensor$(2^{++} )$ state, which is formed from a color anti-triplet diquark and a color triplet anti-diquark, the process involves the creation of a $c\bar{c}$ pair connected by a single gluon to the spectator quarks. For the scalar$(0^{++} )$ state, one must consider contributions from both the color anti-triplet$(\bar{3}\otimes3 )$ and color sextet$(6\otimes\bar{6})$ diquark-antidiquark configurations. The resulting scattering amplitudes are evaluated to the lowest order in the relative velocity of the heavy quarks, and by comparing with the operator matrix elements on the NRQCD side, we extract the corresponding SDCs.

After evaluating the relevant color and spinor traces using well-known projector techniques for each contributing diagram, we arrive at the LO results for the coefficients

\begin{eqnarray}
    \tilde F_{\rm TT,0} = \frac{4 \pi \alpha_s e_Q^2}{\sqrt{3}m_Q^4}, \quad \tilde F_{\rm TT,2} = \frac{128 \pi \alpha_s e_Q^2}{\sqrt{3}m_Q^2},  \,  
\end{eqnarray}
and
\begin{eqnarray}
 F^{\overline{\bf 6} {\bf 6}}_{
   \rm TT} =  - \frac{8 \sqrt{2} \pi \alpha_s e_Q^2}{\sqrt{3} m_Q^2}, \quad
    F^{\overline{\bf 3} {\bf 3}}_{
   \rm TT} = \frac{48 \pi \alpha_s e_Q^2}{m_Q^2} . 
\end{eqnarray}
Our SDCs for the $0^{++}$ and $J_z = \pm 2$ component of the $2^{++}$ state, when squared, agree with the ones obtained earlier in Ref.\cite{Sang:2023ncm}.

We now obtain  the decay width of the tensor tetraquark:
\begin{eqnarray}
    \Gamma(T_{4Q}(2^{++}) \to \gamma \gamma) &=&   \frac{\pi \alpha_{\rm em}^2}{5 M} |F_{\rm TT,2}|^2( 1 + R_0) \nonumber \\
    &=& \frac{128 \pi^3 \alpha_{\rm em}^2 \alpha_s^2 e_Q^4}{15 m_Q^8} \Big\{ 1 + \frac{1}{6} \Big( \frac{M}{4m_Q} \Big)^4\Big\} \, \Big| \bra{0}| {\cal O}^{(2)}_{\bf \bar 3 \otimes 3} |\ket{T_{4Q}} \Big|^2 .  
\end{eqnarray}
Here, the relative contribution of the $J_z=0$ polarization to the decay is given by
\begin{equation}
    R_0 = \frac{1}{6} \Big( \frac{M}{4m_Q} \Big)^4 .
\end{equation}
The operator matrix element squared has dimension GeV$^9$ and is related to the wavefunction at the origin, such that
\begin{eqnarray}
    \Big| \bra{0}| {\cal O}^{(2)}_{\bf \bar 3 \otimes 3} |\ket{T_{4Q}} \Big|^2 = 16 \, |\psi(0)|^2. 
\end{eqnarray}
Due to the mixing of color configurations, for the scalar tetraquark, the result is more involved and reads
\begin{eqnarray}
      \Gamma(T_{4Q}(0^{++}) \to \gamma \gamma) &=& \frac{\pi^3 \alpha_{\rm em}^2 \alpha_s^2 e_Q^4}{m_Q^8} \Big\{ 18  \Big| \bra{0}| {\cal O}^{(0)}_{\bf \bar 3 \otimes 3} |\ket{T_{4Q}} \Big|^2 + \frac{1}{3}\Big| \bra{0}| {\cal O}^{(0)}_{\bf 6 \otimes \bar 6} |\ket{T_{4Q}} \Big|^2 \nonumber \\
      &-& 2 \sqrt{6} \Re e \Big( \bra{0}| {\cal O}^{(0)}_{\bf \bar 3 \otimes 3} |\ket{T_{4Q}} \bra{0}| {\cal O}^{(0)}_{\bf 6 \otimes \bar 6} |\ket{T_{4Q}}^*\Big) \Big\}. 
\end{eqnarray}
Below, we will also use the following shorthand notation for the square of the LDMEs:
\begin{eqnarray}
\ave {{\cal O}_{ \bar {\bf R} \otimes {\bf R}}^{(J)}} = \Big| \bra{0}| {\cal O}^{(J)}_{\bf \bar R \otimes R} |\ket{T_{4Q}} \Big|^2 , \quad \ave{{\cal O}_{\rm mix}^{(0)}} = \Re e \Big( \bra{0}| {\cal O}^{(0)}_{\bf \bar 3 \otimes 3} |\ket{T_{4Q}} \bra{0}| {\cal O}^{(0)}_{\bf 6 \otimes \bar 6} |\ket{T_{4Q}}^*\Big) .
\end{eqnarray}

\section{Quark models for the fully heavy compact tetraquark}
\label{section:LDMEs}

In order to estimate the LDMEs necessary for the calculation of decay widths, we have to resort to models of the $T_{4c}$ states. In this work we concentrate on the description of these states as compact $QQ \bar Q \bar Q$ systems. In the following we use four-body wave functions obtained in Ref. \cite{Lu:2020cns}.

The model employed in these references is an extended relativized quark model, which provides a robust framework for studying the spectroscopy and structure of multi-quark hadrons. The core of the model is the numerical solution of the four-body Schr\"odinger equation for the $cc\bar{c}\bar{c}$ system. The Hamiltonian includes relativistic kinetic energy terms for each quark and a potential energy term describing the interactions between all pairs of constituents
\begin{eqnarray}
H = \sum_{i=1}^4 \sqrt{\mathbf{p}_i^2 + m_i^2} + \sum_{i<j} V_{ij}(\mathbf{r}_{ij}),
\end{eqnarray}
where the two-body potential $V_{ij}$ encodes the essential infrared features of QCD: a linear confining term and a one-gluon-exchange (OGE) interaction that includes the spin--spin, spin--orbit, and tensor forces. Each term is dressed by the color factor $\mathbf{F}_i\cdot\mathbf{F}_j$ of the interacting quark pair, ensuring consistency with the color algebra of the four-quark system. The four-body problem is solved with high precision using the Gaussian Expansion Method (GEM)~\cite{Hiyama:2003cu}, a variational technique specifically designed for few-body systems.

Within GEM, the spatial part of the $S$-wave tetraquark wave function is expanded in a set of Gaussian basis functions distributed over the three independent Jacobi coordinates $\{\mathbf{r}_{12},\,\mathbf{r}_{34},\,\mathbf{r}\}$ 
(see Fig.~1 of Ref.~\cite{Lu:2020cns}):
\begin{eqnarray}
\Psi(\mathbf{r}_{12}, \mathbf{r}_{34}, \mathbf{r}) \; &=& \; 
\sum_{n_{12},\,n_{34},\,n} C_{n_{12}n_{34}n}\;
\psi_{n_{12}}(\mathbf{r}_{12})\,\psi_{n_{34}}(\mathbf{r}_{34})\,\psi_{n}(\mathbf{r}),
\\[4pt]
\psi_{n}(\mathbf{r}) \; &=& \; 
\left(\frac{2\nu_n}{\pi}\right)^{\!3/4} e^{-\nu_n r^{2}} \,Y_{00}(\hat{\mathbf{r}}),
\end{eqnarray}
where the Gaussian width parameters follow a geometric progression,
\begin{eqnarray}
\nu_n = \frac{1}{r_{1}^{2}\,a^{2(n-1)}}, \quad n = 1,\dots,N_{\max},
\end{eqnarray}
a choice that guarantees dense coverage of the short-range region and an accurate description of long-range tails simultaneously~\cite{Hiyama:2003cu}. The expansion coefficients $C_{n_{12}n_{34}n}$ and the tetraquark mass are obtained by solving the generalized eigenvalue problem
\begin{eqnarray}
(H_{ij} - E\,\mathcal{N}_{ij})\,\mathbf{C}_j = 0,
\end{eqnarray}
with the $\mathcal{N}_{ij}$ representing the overlap matrix arising from the non-orthogonality of the Gaussian bases.

A central quantity for quarkonium-like production and decay phenomenology is the wave function at the origin,
\begin{eqnarray}
\psi(0) \equiv \Psi(\mathbf{0},\mathbf{0},\mathbf{0})
= \sum_{n_{12},\,n_{34},\,n} C_{n_{12}n_{34}n}
\left(\frac{2\nu_{n_{12}}}{\pi}\right)^{\!3/4}
\left(\frac{2\nu_{n_{34}}}{\pi}\right)^{\!3/4}
\left(\frac{2\nu_{n}}{\pi}\right)^{\!3/4},
\end{eqnarray}
since the LDMEs of fully-heavy tetraquarks are proportional to $|\psi(0)|^{2}$. Using the numerical solutions and Gaussian parameters reported in Ref.~\cite{Lu:2020cns}, we have extracted $\psi(0)$ for each independent color configuration of both the ground $(1S)$ and first radially excited $(2S)$ states.

For scalar ($0^{++}$) tetraquarks, the physical mass eigenstates are linear combinations of the color-antitriplet ($\bar{3}\otimes 3$) and color-sextet ($6\otimes\bar{6}$) diquark-antidiquark configurations. 
One obtains two physical states that are mixtures of the two color configurations
\begin{eqnarray}
 \begin{pmatrix}
 \psi_{0^{++}} \\
 \psi'_{0^{++}}
 \end{pmatrix}  =
 \begin{pmatrix}
 \cos \theta && - \sin \theta \\
 \sin \theta && \cos \theta
\end{pmatrix}
\begin{pmatrix}
\psi_{\bar 3 3} \\
\psi_{6 \bar 6} 
\end{pmatrix}.
\end{eqnarray}
For the spin-$0$ states this configuration mixing is not small, and must be taken into account in the calculation of the LDMEs, which split up according to
\begin{eqnarray}
    \ave {{\cal O}_{ \bar {\bf 3} \otimes {\bf 3}}^{(0)}}
= 
\begin{cases}
  16 \cos^2 \theta \, |\psi_{\bar 3 3}(0)|^2 \\
  16 \sin^2 \theta \, |\psi_{\bar 3 3}(0)|^2 
\end{cases} \, , \quad 
    \ave {{\cal O}_{ {\bf 6} \otimes \bar {\bf 6}}^{(0)}}
= 
\begin{cases}
 16 \sin^2 \theta \, |\psi_{6 \bar 6}(0)|^2 \\
 16 \cos^2 \theta \, |\psi_{6  \bar 6}(0)|^2
\end{cases}.
\label{eq:LDME_spin0}
\end{eqnarray}
and
\begin{eqnarray}
\ave {{\cal O}_{\rm mix}^{(0)}}
= 
\begin{cases}
  - 16 \cos \theta  \sin \theta \, \psi_{\bar 3 3}(0) \psi_{6 \bar 6}(0) \\
  16 \cos \theta \sin \theta  \, \psi_{\bar 3 3}(0) \psi_{6 \bar 6}(0) 
\end{cases}
\label{eq:LDME_mix}
\end{eqnarray}
For tensor ($2^{++}$) states the sextet-anti-sextet configuration is not possible in the $S$--wave and thus is expected to lie significantly higher in energy and is neglected; these states are treated as pure color-antitriplet-triplet systems.

The resulting values of $\psi(0)$, which serve as the primary non-perturbative input for our subsequent decay-width predictions, are collected in Table~\ref{table:wavefunctions}. In the table, we also show the wave function at the origin of the relevant physical state, but we caution the reader, that for the calculation of the decay width, the LDMEs of eqn.~(\ref{eq:LDME_spin0},\ref{eq:LDME_mix}) enter. The dependence on the state is somewhat counterintuitive: differently from what we are used to from the two--body state, $\psi(0)$ of the $2s$ states is  consistently larger than the one of $1s$ states. For the $3s$ states then, $\psi(0)$ vanishes within the accuracy of the calculation. 
\begin{table}[htb]
\centering
\caption{Calculated wave functions at the origin for various candidate tetraquark states. Units are MeV for mass and GeV$^{\frac{9}{2}}$ for wave functions.}
\vspace{0.3 cm}
\label{table:wavefunctions}
\begin{ruledtabular}
\begin{tabular}{l l l c c c}
Mass ($J^{PC}$) & State & Eigenvector & $\psi_{\bar{3}3}(0)$ & $\psi_{6\bar{6}}(0)$ & $\psi_{\text{phys}}(0)$\\
\hline
6435 ($0^{++}$) & 1s & [0.62, 0.79] & 0.0466 & 0.0283 & 0.0510 \\
6542 ($0^{++}$) & 1s & [0.79, -0.62] & 0.0466 & 0.0283 & 0.0189 \\
\addlinespace[4pt]
6849 ($0^{++}$) & 2s & [0.50, 0.87] & 0.0721 & 0.0466 & 0.0764 \\
6940 ($0^{++}$) & 2s & [0.87, -0.50] & 0.0721 & 0.0466 & 0.0391\\
\addlinespace[4pt]
7025 ($0^{++}$) & 3s & [0.664, 0.748] & $-2.69\times10^{-9}$ &$-2.67\times10^{-9}$ &$-3.78\times10^{-9}$ \\
7063 ($0^{++}$) & 3s & [0.748, -0.664] & $-2.69\times10^{-9}$ &$-2.67\times10^{-9}$ &$-2.40\times10^{-10}$ \\ 
\hline
6543 ($2^{++}$) & 1s & 1 & 0.0300 & --- & 0.0300\\
6928 ($2^{++}$) & 2s & 1 & 0.0471 & --- & 0.0471\\
7064 ($2^{++}$) & 3s & 1 & $-2.72\times10^{-9}$ & --- &$-2.72\times10^{-9}$\\
\end{tabular}
\end{ruledtabular}
\label{tab:WFs}
\end{table}

Regarding the strong coupling constant, $\alpha_s$ is evolved at two-loop accuracy using the \textsc{RunDec} package~\cite{Chetyrkin:2000yt}. Starting from the world-average value $\alpha_s^{(5)}(m_Z) = 0.1179$, we cross the flavour thresholds sequentially: $\alpha_s^{(5)}(m_Z) \to \alpha_s^{(5)}(m_b) \to \alpha_s^{(4)}(m_b) \to \alpha_s^{(4)}(m_c) \to \alpha_s^{(3)}(m_c)$, with $m_b = 4.75\,\rm{GeV}$ and $m_c = 1.50\,\rm{GeV}$. At the characteristic scale of the tetraquark, $\mu \simeq 2m_c$, this procedure yields $\alpha_s^{(3)}(3\,\mathrm{GeV}) \simeq 0.24$. Combining the perturbatively calculated SDCs with the non-perturbative LDMEs extracted from the wave functions at origin and the strong coupling constant, we arrive at our predictions for the two-photon decay widths. The full set of results, including the physical LDMEs for each color configuration, the root-mean-square radii characterising the spatial extent of each state, and the predicted decay widths $\Gamma_{\gamma\gamma}$ with their associated scale-variation uncertainties, is collected in Table~\ref{table:LDME}. We have defined the mixing terms as the interference contributions
$\langle 0|\mathcal{O}^{(0)}_{\bar{3}\otimes 3}|T_{4Q}\rangle\,\langle 0|\mathcal{O}^{(0)}_{6\otimes\bar{6}}|T_{4Q}\rangle^{*}$,
each multiplied by the respective eigenvector components. 

For the compact tetraquark scenario, the LDMEs are obtained from the (squares of) the respective wave functions at the origin. In the case of the scalar states, we further resolve the LDMEs into their $\bar{3}3$, $6\bar{6}$, and interference components, taking account of the mixing between these components. These enter the NRQCD factorization formula with different short-distance weights.

The CMS experiment prefers the $2^{++}$ assignment for the observed states. It seems therefore natural to identify the $(2^{++}, 1s)$ state with $X(6600)$, the $(2^{++},2s)$ state with the $X(6900)$ and finally the $(2^{++},3s)$ state with the $X(7100)$. The $X(6400)$, reported by the ATLAS collaboration, finds a possible candidate in the $(0^{++},1s)$ state. As is evident from Table~\ref{table:LDME}, the states considered by us span a range of two-photon decay widths from approximately $0.1$ to $0.69\,\rm{keV}$, with the exact value depending sensitively on the color mixing pattern and the total spin of the state. The corresponding RMS radii, ranging from $0.35$ to $0.52\,\rm{fm}$, confirm the compact nature of these tetraquark configurations.

\begin{table}
\centering
\caption{Predictions for masses (MeV), root of mean square (RMS) radii (fm), LDMEs (GeV$^{9}$), and two-photon decay widths (keV). The masses and RMS radii agree with Ref.~\cite{Lu:2020cns}. Uncertainties on $\Gamma_{\gamma\gamma}$ are dominated by scale variation and truncations of higher-order renormalization evolution.}
\label{table:LDME}
\begin{ruledtabular}
\begin{tabular}{l l c c c c c c}
\multirow{2}{*}{Mass ($J^{PC}$)} & \multirow{2}{*}{State} & \multirow{2}{*}{Eigenvec.} & \multirow{2}{*}{$\langle r^2 \rangle^{\frac{1}{2}}$ (fm)} & \multicolumn{3}{c}{LDMEs} & \multirow{2}{*}{$\Gamma_{\gamma\gamma}$ (keV)} \\
\cmidrule(lr){5-7}
& & & &  $\ave{{\cal{O}}_{\bf \bar 3 \bf 3}^{(J)}}$ & $\ave{{\cal{O}}_{\bf 6 \bf \bar 6}^{(J)}}$  & $\ave{{\cal{O}}_{\rm mix}^{(J)}}$ &\\
\hline
6435 ($0^{++}$) & 1s & [0.62, 0.79] & 0.265 & 0.01323 & 0.00794 & \phantom{-}0.010 & ${0.106^{+0.03}_{-0.02}}$ \\
6542 ($0^{++}$) & 1s & [0.79,-0.62] & 0.281 & 0.02152 & 0.00488 & -0.010 & ${0.245^{+0.07}_{-0.05}}$ \\
\addlinespace[4pt]
{6849 ($0^{++}$)} & {2s} & {[0.5, 0.87]} & {0.351} & {0.02078} & {0.02605} & {0.023} & ${0.149^{+0.05}_{-0.03}}$ \\
{6940 ($0^{++}$)} & {2s} & {[0.87,-0.50]} & {0.452} & {0.06234} & {0.00868} & {-0.023} & ${0.688^{+0.21}_{-0.15}}$ \\
\hline
6543 ($2^{++}$) & 1s & 1 & 0.321 & 0.01442 & --- & --- & ${0.088^{+0.03}_{-0.02}}$ \\
{6928 ($2^{++}$)} & {2s} &{1} & {0.520} & {0.03546} & --- & --- & ${0.217^{+0.07}_{-0.05}}$ \\
\end{tabular}
\end{ruledtabular}
\end{table}

If we identify the $(2^{++}, 1s)$ state with the $X(6600)$ and the $(2^{++},2s)$ state with the $X(6900)$, we obtain the $\gamma \gamma$ branching ratios of these resonances as:
\begin{eqnarray}
    {\rm Br}(X(6600) \to \gamma \gamma) &=& (1.97^{+0.67}_{-0.45})\times 10^{-7},\\
    {\rm Br}(X(6900) \to \gamma \gamma) &= &(1.61^{+0.52}_{-0.37})\times 10^{-6}.
\end{eqnarray}
These are obtained directly from the total widths measured in \cite{CMS:2026tiu}: $\Gamma_{\rm tot}(X(6600)) = 0.446\,\rm{GeV}$ and $\Gamma_{\rm tot}(X(6900)) = 0.135\,\rm{GeV}$, as shown in Table \ref{tab:T4c_states}, together with the $\Gamma_{\gamma\gamma}$ values from Table~\ref{table:LDME}.
It is interesting to note that for the $X(7100)$ candidate, the $(2^{++},3s)$ state, the model predicts a vanishing two--photon decay width.

A brief comparison with earlier results from the literature, which span a very large range, is in order. Our results for the two--photon width are about two orders of magnitude larger than the earliest quark model estimates for the spin-0 $C$-even tetraquark from Ref.\cite{Badalian:1985es}, who estimated $\Gamma(T_{4c} \to \gamma \gamma) \approx 10^{-4} \,  \Gamma(\eta_c \to \gamma \gamma) \sim 1 {\rm eV}$. In Ref. \cite{Goncalves:2021ytq}, the tetraquark two--photon width is assumed to be equal to the one of the quarkonium with the same quantum numbers, and it is thus proposed that $\Gamma(T_{4c} \to \gamma \gamma) \sim 0.5 \div 2 \, {\rm keV}$.

More recently, the two--photon decay width was estimated in Ref. \cite{Esposito:2021ptx} within a vector meson dominance (VDM) approach, relating it to the branching fraction to $J/\psi J/\psi$. This paper obtains, for example for the $X(6900)$ tensor state, that
\begin{eqnarray}
 \Gamma (X(6900) \to \gamma \gamma) = 0.086 \, {\rm keV} \, \times {\rm Br}(X(6900) \to J/\psi J/\psi).
\end{eqnarray}
This is somewhat smaller, but in a similar ballpark as our results, if we assume that ${\rm Br}(X(6900) \to J/\psi J/\psi) \sim 1$. We will come back to the issue of the $J/\psi J/\psi$ branching in Section \ref{sec:UPC_psipsi}.

Among the results closest to ours are the NRQCD calculations performed at leading order in Ref.~\cite{Sang:2023ncm} and at next-to-leading order in Ref.~\cite{Liu:2025mxv}. The "Model I" adopted in those studies is based on the same extended relativized quark model of Ref.~\cite{Lu:2020cns} that we employ. Those authors present results for one scalar and one tensor state. The tensor state appears to be the $1s$ resonance, which we identify as $X(6600)$, although it is labelled $X(6900)$ in Ref.~\cite{Sang:2023ncm}. Results for the radially excited $2s$ states are not provided. For the scalar sector, it is not fully transparent to us whether the color-configuration mixing, which gives rise to two physical eigenstates with the mixing coefficients listed in Table~\ref{tab:WFs}, has been taken into account in  Ref.~\cite{Sang:2023ncm}. Despite these differences in scope, the final numerical results lie in a similar ballpark as ours. The NLO corrections to the strong coupling computed in Ref.~\cite{Liu:2025mxv} are found to be rather mild.

Exceptionally large $\gamma \gamma$ decay widths are obtained in  Refs.\cite{Biloshytskyi:2022dmo,Kalamidas:2025gen}.
In Ref.\cite{Biloshytskyi:2022dmo}, an attempt was made to resolve a certain discrepancy of ATLAS light--by--light (LbL) scattering data \cite{ATLAS:2020hii} with theoretical predictions by exhausting the missing cross section with a tetraquark contribution. Only a spin-$0$ tetraquark is considered, and values of $\Gamma(X(6900) \to \gamma \gamma) \sim 45 \div 67 \, \rm{keV}$, about two orders of magnitude larger than our results, are obtained from a fit to LbL data. These authors also quote $\gamma \gamma$ branching fractions obtained from a VDM ansatz, ${\rm Br}(X(6900) \to \gamma \gamma) \sim 3 \times 10^{-6}$, but this result is again contingent on the assumption that ${\rm Br}(X(6900) \to J/\psi J/\psi) \sim {\cal O}(1)$.

Finally, Ref.\cite{Kalamidas:2025gen} adopts an effective diquark--antidiquark two--body model of tetraquarks with wave functions obtained in Ref.\cite{Debastiani:2017msn}. For the $^1S_0$ ground state $\Gamma(T_{4c} \to \gamma \gamma) \sim 82.1 \, {\rm keV}$ is obtained, whereas for the first and second $^5S_2$ states $\Gamma(T_{4c} \to \gamma \gamma) \sim 14.7 \, {\rm keV}$ and $\Gamma(T_{4c} \to \gamma \gamma) \sim 5.1 \, {\rm keV}$ respectively are given. In the point-like diquark (PLD) effective theory of Ref.~\cite{Kalamidas:2025gen}, the $cc$ pair is treated as an elementary spin-$1$ color-antitriplet field. The tetraquark reduces to an effective two-body $(\bar{3}\otimes 3)$ bound state, with the internal structure of the diquark subsumed into a single local matrix element rather than resolved through explicit few-body dynamics. This simplification tends to overestimate the wave function at the origin. By contrast, the four-body GEM calculation~\cite{Lu:2020cns} employed in the present work accounts for the full $cc\bar{c}\bar{c}$ dynamics: in the spin--0 case, the mixing with the $6\otimes\bar{6}$ color configuration and the finite spatial extent of each diquark cluster both act to suppress the predicted LDMEs.
In the PLD framework, the analogue of our SDC is the annihilation of pointlike spin-1 diquark and antidiquark into photons.
We trace part of the difference between the very large numbers in comparison to our results to this oversimplified treatment of the annihilation process.

\section{$T_{4c}$ production in UPCs}
\label{sec:UPC}
We now turn to the production of $T_{4c}$ states in ultraperipheral heavy ion collisions (UPCs) (see e.g. reviews \cite{Baur:2001jj,Schafer:2020bnm}).
\begin{eqnarray}
\sigma(AA \to T_{4c} AA) = \int dW^2 \, \frac{d{\cal{L}}_{\gamma \gamma} (W^2)}{dW^2} \sigma(\gamma \gamma \to T_{4c}; W^2),    
\end{eqnarray}
where the effective photon--photon luminosity is given in terms of the Weizs\"acker--Williams fluxes of photons as 
\begin{eqnarray}
  \frac{d{\cal{L}}_{\gamma \gamma}(W^2)}{dW^2}  &=& \int \frac{dx_1}{x_1} \frac{dx_2}{x_2} \delta(W^2 - x_1 x_2 s_{\rm NN}) \int d^2\bb_1 d^2\bb_2 \,
    S^2(|\bb_1 - \bb_2|) N(x_1,\bb_1) N(x_2,\bb_2), \nonumber \\
\end{eqnarray}
where for our purposes the photon fluxes for pointlike sources are sufficient:
\begin{eqnarray}
 x N_A(x, \bb) = \frac{Z^2 \alpha_{\rm em}}{\pi} \frac{1}{b^2} \, (x m_N b)^2 K_1^2(xm_N b)\, .   
\end{eqnarray}
Here $m_N$ is the nucleon mass, and $Z=82$ for the lead nucleus. For the survival factor we use $S^2(|\bb|) = \theta(2 R_A - |\bb|)$, with $R_A \sim 7 \, {\rm fm}$. The total cross section of the production of the Spin-$J$ resonance is given by a standard Breit--Wigner formula
\begin{eqnarray}
 \sigma( \gamma \gamma \to T_{4c}(J);W) = 8 \pi (2J+1) \frac{M_{T_{4c}}}{W} \, \frac{ \Gamma_{\rm tot} \Gamma_{\gamma \gamma}}{(W^2 - M_{T_{4c}}^2)^2 + \Gamma_{\rm tot}^2 M_{T_{4c}}^2} \, . 
\end{eqnarray}
In Table~\ref{table:LDME} we present our results for the two-photon decay widths of the $S$-wave spin-0 and spin-2 resonances considered in this work, together with the corresponding $\gamma\gamma$ branching fractions. For definiteness we assume a total width $\Gamma_{\rm tot}=0.1\,\mathrm{GeV}$ for all states~\cite{Sang:2023ncm} except $\Gamma_{\rm tot}=0.446\,\rm{GeV}$ for the $6543\,\rm{MeV}$ and $\Gamma_{\rm tot}=0.135\,\rm{GeV}$ for the $6928\,\rm{MeV}$~\cite{CMS:2026tiu}.

From these decay widths we obtain the resonant peak cross sections for $\gamma\gamma \to T_{4c}$ in ultraperipheral Pb--Pb collisions at $\sqrt{s_{NN}} = 5.5\,\mathrm{TeV}$, evaluated using the wave functions of Ref.~\cite{Lu:2020cns}. The results are collected in Table~\ref{table:peak_cross_sections}.
\begin{table}[htb]
\centering
\caption{
Peak cross sections for $\gamma\gamma \to T_{4c}$ and total resonance production cross section for $AA \to AA\,T_{4c}$ for 
$A = {^{208}Pb}$ and $\sqrt{s_{\rm NN}} = 5.5 \, {\rm TeV}$. Total widths are $\Gamma_{\rm tot}=0.446\,\rm{GeV}$ for the $6543\,\rm{MeV}$ state, $\Gamma_{\rm tot}=0.135\,\rm{GeV}$ for the $6928\,\rm{MeV}$ state, and $\Gamma_{\rm tot}=0.1\,\rm{GeV}$ for all other states.
}
\label{table:peak_cross_sections}
\begin{ruledtabular}
\begin{tabular}{llcc}
Mass ($J^{PC}$) & State & $\sigma(\gamma\gamma\to T_{4c})$ (nb) & $\sigma(AA\to T_{4c} AA)$($\mu$b) \\
\hline
6435 ($0^{++}$) & 1s & $0.251$&$0.351$ \\
6542 ($0^{++}$) & 1s & $0.560$&$0.769$\\
\addlinespace
6849 ($0^{++}$) & 2s & $0.311$&$0.396$ \\
6940 ($0^{++}$) & 2s & $1.398$&$1.738$ \\
\hline
6543 ($2^{++}$) & 1s & $1.006$&$1.164$ \\
6928 ($2^{++}$) & 2s & $2.212$&$2.733$ \\
\end{tabular}
\end{ruledtabular}
\end{table}

A notable feature of Table~\ref{table:peak_cross_sections} is the hierarchy among scalar states: the eigenstates with destructive color interference in $\psi_{\mathrm{phys}}(0)$ ($6542$ and $6940\,\rm{MeV}$) yield larger cross sections than their constructive counterparts ($6435$ and $6849\,\rm{MeV}$). This inversion arises because the mixing term $\langle\mathcal{O}_{\bar{3}3}\rangle\langle\mathcal{O}_{6\bar{6}}\rangle^{*}$ enters $\Gamma_{\gamma\gamma}$ with a sign fixed by the relative phase of the color-dependent SDCs. The tensor states profit from the $(2J+1)=5$ spin factor and are systematically the most abundantly produced, with the $2s$ candidate at $6928\,\rm{MeV}$ reaching $\sigma = 2.2\,\mathrm{nb}$. These cross sections are fairly large, corresponding to $2.5\times 10^{5}$--$2.2\times 10^{6}$ events per $\mathrm{fb}^{-1}$. 

Also shown in Table~\ref{table:peak_cross_sections} are total cross sections for resonance production in $^{208}Pb + {^{208}Pb}$ UPCs at LHC energies. They are on the order of microbarns, which must be compared to typical luminosities of a few nb$^{-1}$.
The experimental feasibility, however, depends crucially on the accessible decay channels and the level of irreducible background in each final state.

In this paper we will first discuss the $J/\psi J/\psi$ final state, which has been the discovery channel of the fully charmed tetraquarks. We then revisit the possible impact of tetraquark states on light--by--light(LbL) scattering.

\subsection{$T_{4c}$ and production of $J/
\psi J/\psi$ pairs in UPC}
\label{sec:UPC_psipsi}

We calculate the cross section for the process $AA \to J/\psi J/\psi AA$ in UPCs from
\begin{eqnarray}
    \frac{ d \sigma}{dy_1 dy_2 d\bp_\perp^2} &=& \int d^2\bb_1 d^2\bb_2  \, S^2(|\bb_1 - \bb_2|) N(x_1,\bb_1) N(x_2,\bb_2) \,  \frac{d \hat \sigma(\gamma \gamma \to J/\psi J/\psi)}{d \hat t}\,, 
\label{eq:upc_master}
\end{eqnarray}
where
\begin{eqnarray}
    x_1 = \frac{m_\perp}{
    \sqrt{s_{    \rm NN}}}\Big( e^{y_1} + e^{y_2} \Big) \, , \quad x_2 = \frac{m_\perp}{
    \sqrt{s_{    \rm NN}}}\Big( e^{-y_1} + e^{-y_2} \Big) \, , \quad m_\perp = 
    \sqrt{\bp_\perp^2 + m_{J/\psi}^2} . 
\end{eqnarray}
Here $\sqrt{s_{\rm NN}}$ is the per--nucleon energy in the $AA$ center of mass, and the Mandelstam variables for the process $\gamma \gamma \to J/\psi J/\psi$ are
\begin{eqnarray}
    \hat s &=& M^2_{J/\psi J/\psi} = x_1 x_2 s_{\rm NN} = 2m_\perp^2 ( 1 + \cosh (y_1 - y_2) ), \\
    \hat t &=& - m_\perp^2( 1 + e^{(y_1- y_2)}) + m_{J/\psi}^2, \quad  \hat u = - m_\perp^2( 1 + e^{(y_2- y_1)}) + m_{J/\psi}^2.
\end{eqnarray}
In the absence of detailed knowledge on the coupling of the individual tetraquark states to $J/\psi J/\psi$, we follow Ref.\cite{Esposito:2021ptx} and adopt the simplest structure of the $T_{4c} \to J/\psi J/\psi$ vertex independent of the decay momentum.

The matrix elements for the decay of a scalar or tensor exotic meson into two vector mesons
are given as
\begin{eqnarray}
\langle T_{4c}(0^{++})|V_1 V_2 \rangle
&=& \alpha_0\,\varepsilon_1\!\cdot\!\varepsilon_2
\,,\\
\langle T_{4c}(2^{++})|V_1 V_2 \rangle
&=& \alpha_2 \, E_{\mu\nu}\,\,\varepsilon_1^\mu\varepsilon_2^\nu\, .
\end{eqnarray}
Here $\varepsilon_i$ are the polarization vectors of the two vector mesons, $\alpha_{0,2}$ are couplings, and $E_{\mu\nu}$ is the polarization tensor of the spin-2 state. The effective couplings $|\alpha_{0,2}|^2$ are obtained from the input partial widths via
\begin{eqnarray}
\Gamma\bigl(T_{4c}(0^{++})\to J/\psi J/\psi\bigr)
&=& 
\frac{|\alpha_0|^2\,p}{16\pi M_{0^{++}}^2}\,
\left(3-\frac{ M_{0^{++}}^2}{m_\psi^2} + \frac{1}{4} \frac{ M_{0^{++}}^4}{m_\psi^4} \right), 
\nonumber\\
\Gamma\bigl(T_{4c} (2^{++})\to J/\psi J/\psi\bigr)
&=& 
\frac{|\alpha_2|^2\,p}{16\pi M_{2^{++}}^2}\,
\left(\frac{7}{15} + \frac{1}{10}\frac{M_{2^{++}}^2}{m_\psi^2} + \frac{1}{120}\frac{M_{2^{++}}^4}{m_\psi^4}\right)\,, 
\end{eqnarray}
where $p=\sqrt{M^2/4-m_\psi^2}$ denotes the decay momentum at the resonance peak. 

While experiments have determined the total decay width of the observed states (see Table~\ref{tab:T4c_states}), the branching fractions to $J/\psi J/\psi$ are not known.
Many of the earlier works, e.g. \cite{Esposito:2021ptx,Biloshytskyi:2022dmo}, assume $B_\psi \equiv {\rm Br} (T_{4c} \to J/\psi J/\psi) \approx {\cal O}(1)$.
This is in stark contrast to Ref. \cite{Becchi:2020uvq}, who argue that ${\rm Br} (T_{4c}(2^{++}) \to \mu^+ \mu^- \mu^+ \mu^-) \sim 10^{-5}$, and ${\rm Br}(T_{4c}(2^{++}) \to J/\psi \mu^+ \mu^-) \sim 1.7 \times 10^{-4}$ so that using ${\rm Br}(J/\psi \to \mu^+ \mu^-) \approx 0.06$, one would estimate $B_\psi \sim 3 \times 10^{-3}$.
Such rather small branching fractions appear to be also required by estimates of the measured event rates in $pp$ scattering \cite{Belov:2024qyi,Wang:2025hex}. Finally we wish to point out, that model calculations with large branching fractions -- $B_\psi \sim {\cal{O}}(10 \%)$ or larger -- are also present in the literature, see e.g. \cite{Lu:2025lyu}.

For definiteness, we shall assume a small branching ratio $B_\psi = 0.003$. We use the same value for all resonances, although we are aware of the fact that $B_\psi$ may strongly depend on the state, see for example \cite{liu:2020eha}.

We use the masses $M_{0^{++}} = 6.940\,\rm{GeV}$, $M_{2^{++}} = 6.928\,\rm{GeV}$, a total width $\Gamma_{\rm{tot}} = 0.135\,\rm{GeV}$ for $6.928\,\rm{GeV}$ while $\Gamma_{\rm{tot}} = 0.446\,\rm{GeV}$ for $6.593\,\rm{GeV}$~\cite{CMS:2026tiu} and $m_\psi = 3.096\,\rm{GeV}$. We then obtain the coupling constants
\begin{eqnarray}
|\alpha_0|^2(6940)&\simeq& 8.66 \times 10^{-2}\,\rm{GeV},\\
\nonumber
|\alpha_2|^2(6928)&\simeq& 4.561 \times 10^{-1}\,\rm{GeV},\\
\nonumber
|\alpha_2|^2(6593)&\simeq& 1.870\,\rm{GeV}.
\end{eqnarray}
With this minimal coupling, the differential cross section for the scalar resonance $\gamma\gamma \to T_{4c}(0^{++}) \to J/\psi J/\psi$ reads

\begin{eqnarray}
\frac{d \sigma (0^{++})}{d \hat t} = \frac{1}{16 \pi \hat s^2} \frac{1}{2!} \frac{1}{4} \frac{e^4 |\alpha_0|^2   F_{\rm TT}^2}{(\hat s - M^2_{0^{++}})^2 + M^2_{0^{++}} \Gamma^2_{\rm tot} } \, 2  \left(3-\frac{ \hat s}{m_\psi^2} + \frac{1}{4} \frac{ \hat s^2}{m_\psi^4} \right),   
\end{eqnarray}
while for the tensor resonance, we have
\begin{eqnarray}
   \frac{d\sigma(2^{++})}{d\that} &=& \frac{1}{16 \pi \hat s^2} \frac{1}{2!} \frac{1}{4} \, \frac{e^4 |\alpha_2|^2}{(\hat s - M^2_{2^{++}})^2 + M^2_{2^{++}} \Gamma_{\rm tot}^2} \, \nonumber \\
   &\times& \Big( 2 F^2_{\rm TT,2} + \frac{4}{3} \hat s^2 F^2_{\rm TT,0} \Big) \Big( \frac{7}{15} + \frac{1}{10} \frac{\hat s} {m^2_{J/\psi}}
   + \frac{1}{120}\frac{\hat s^2}{m^4_{J/\psi}} \Big).
\end{eqnarray}
For narrow resonances, these expressions simplify to
\begin{eqnarray}
    \frac{d \sigma (J^{++})}{d \hat t} &=& 4 \pi (2J+1) \frac{M_J^3}{\hat s^2 p} \frac{\Gamma (T_{4c}(J^{++}) \to \gamma \gamma)  \Gamma (T_{4c}(J^{++}) \to J/\psi J/\psi) }{(\hat s - M^2_{J})^2 + M^2_{J} \Gamma_{\rm tot}^2} \nonumber \\
    &\approx& \frac{4 \pi^2}{M_J^2 p} (2J+1) \frac{\Gamma (T_{4c}(J^{++}) \to \gamma \gamma)  \Gamma (T_{4c}(J^{++}) \to J/\psi J/\psi)}{\Gamma_{\rm tot}} \, \delta(\hat s - M_J^2) \, . 
\end{eqnarray}

These partonic cross sections are inserted as $d\hat{\sigma}/d\hat{t}$ into the impact-parameter--space UPC formula~\eqref{eq:upc_master} and integrated over the two-particle phase space together with the photon fluxes $N(x,\mathbf{b})$ and the nuclear overlap survival factor $S^{2}(\mathbf{b})$. The continuum contribution $\gamma \gamma \to J/\psi J/\psi$ is calculated in LO NRQCD, using the $J/\psi$ wave function at the origin $|\psi(0)|^2= 0.08\,\rm{GeV}^{3}, m_c = 1.50\,\rm{GeV}$ and $\alpha_s = 0.24$.
\begin{figure}[htb]
\includegraphics[width=.48\textwidth]{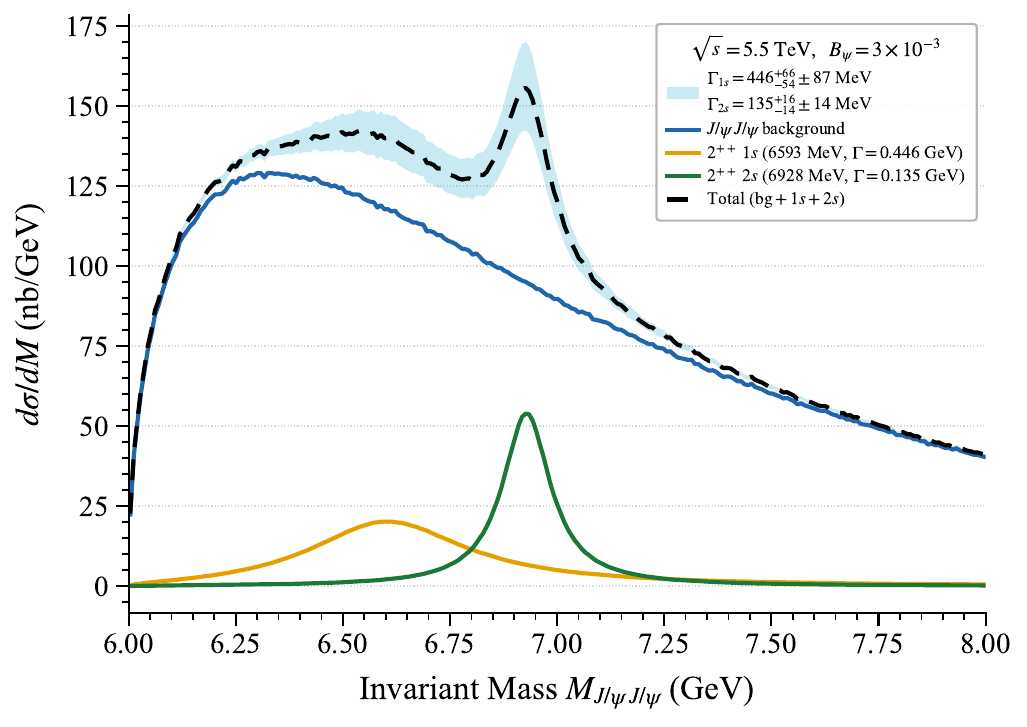}
\includegraphics[width=.48\textwidth]{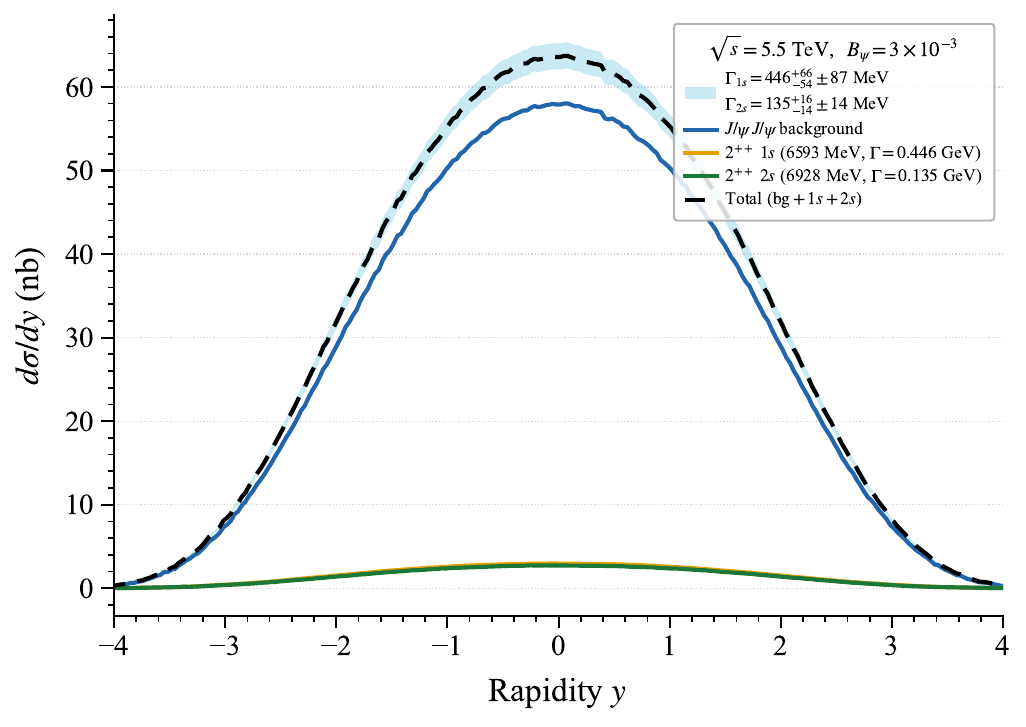}
\caption{
Invariant-mass (left) and rapidity (right) distributions of $J/\psi$-pairs produced in $^{208}\rm{Pb}$\,--\,$^{208}\rm{Pb}$ ultraperipheral collisions at $\sqrt{s_{\rm{NN}}} = 5.5\,\rm{TeV}$. The blue solid curve shows the continuum $\gamma\gamma \to J/\psi J/\psi$ background. 
The rapidity distribution in the right panel was obtained integrating $6 < M_{J/\psi J/\psi} < 20 \, {\rm GeV}$. The orange and green  solid curves correspond to the tensor $2^{++}$ tetraquark resonances: a broad $1s$ state at $6.593\,\rm{GeV}$ ($\Gamma = 0.446\,\rm{GeV}$) and a narrow $2s$ state at $6.928\,\rm{GeV}$ ($\Gamma = 0.135\,\rm{GeV}$), both with widths taken from the CMS di-$J/\psi$ analysis~\cite{CMS:2026tiu}. The black dashed line gives the total $J/\psi J/\psi$ yield including signal and continuum. All curves assume a $J/\psi J/\psi$ branching fraction of $3\times 10^{-3}$.}
\label{fig:dsig_psipsi}
\end{figure}

Figure~\ref{fig:dsig_psipsi} displays the resulting invariant-mass and rapidity distributions. In the invariant-mass spectrum (left panel), the continuum $\gamma\gamma \to J/\psi J/\psi$ (blue solid) constitutes an irreducible background that decreases steeply from $d\sigma/dM \sim 125$\,nb/GeV at $M_{J/\psi J/\psi}=6.25\,\rm{GeV}$ to negligible values above $8\,\rm{GeV}$. Superimposed on this continuum are two tensor $2^{++}$ resonances: a broad $1s$ state at $6.593\,\rm{GeV}$ with $\Gamma = 0.446\,\rm{GeV}$ (orange solid), and a narrow $2s$ state at $6.928\,\rm{GeV}$ with $\Gamma = 0.135\,\rm{GeV}$ (green solid). The widths are taken from the CMS di-$J/\psi$ analysis~\cite{CMS:2026tiu}. The $1s$, despite its large width, produces a substantial enhancement spanning the region $6.3$--$7.0\,\rm{GeV}$, while the $2s$ appears as a sharply localised peak on the upper shoulder of the $1s$ distribution. The black dashed curve shows the total yield, which rises above the continuum across the entire resonance region and peaks near the $2s$ mass. Our calculations do not include possible interferences, neither between resonances, nor between resonances and background.

In the rapidity distribution, both the continuum and the resonant signals are concentrated around $y=0$, as expected for symmetric photon--photon collisions. The rapidity shapes of the two tensor resonances are nearly identical, reflecting the fact that both are produced in the same $\gamma\gamma$ fusion process and their rapidity profiles are governed entirely by the photon fluxes and the survival factor $S^2(b)$, with negligible sensitivity to the resonance mass over the narrow $6-8\,\rm{GeV}$ interval. The combined signal (black dashed) is clearly distinguishable from the continuum alone across the central rapidity region $|y| \lesssim 2$, indicating that with sufficient integrated luminosity and a dedicated low-$p_T$ dimuon trigger, both tensor tetraquark candidates could be studied in the di-$J/\psi$ channel in Pb--Pb UPCs at the HL-LHC.

\subsection{Tetraquark contribution to $\gamma \gamma \to \gamma \gamma$}
LbL scattering in UPCs provides a unique laboratory for precision tests of the Standard Model in a purely electromagnetic final state free from strong-interaction backgrounds. The ATLAS Collaboration has measured the exclusive process $\gamma\gamma \to \gamma\gamma$ in Pb--Pb UPCs at $\sqrt{s_{\rm NN}} = 5.02\,\rm{TeV}$~\cite{ATLAS:2017fur,ATLAS:2020hii}, based on integrated luminosities of $0.48\,\mathrm{nb}^{-1}$ (2015 data) and $1.73\,\mathrm{nb}^{-1}$ (2018 data). The measured diphoton invariant-mass and angular distributions are in good agreement with the pure QED box continuum computed at leading order. On the theoretical side, next-to-leading order (NLO) QCD and N$^3$LO QED corrections to the box continuum have been computed~\cite{AH:2023kor, Bargiela:2026tcn}, the resulting $K$-factors are modest ($\sim$ 5--10\%) in the few-GeV region which makes the continuum prediction a reliably known background for resonance searches.

The resonant cross sections are computed using the same impact-parameter--dependent UPC formalism as for the di-$J/\psi$ channel, with the partonic cross section for a spin-$J$ resonance of mass $M$ given by the Breit--Wigner form. The two-photon partial widths $\Gamma_{\gamma\gamma}$ are taken from Table~\ref{table:LDME}. For the QED box continuum, we employ the leading-order amplitude~\cite{Bern:2001dg}, which is sufficiently accurate for the present purposes given the dominant scale of the continuum.
\begin{figure}[htb]
\includegraphics[width=.8\textwidth]{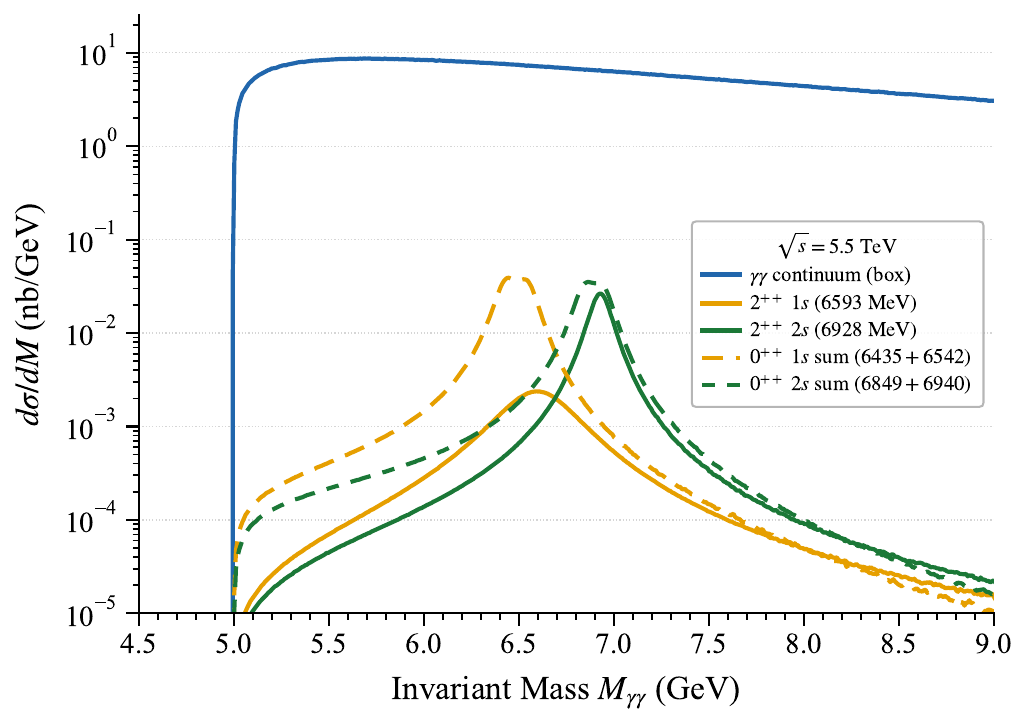}
\caption{
Invariant-mass distribution of the diphoton system produced in $^{208}\mathrm{Pb}$--$^{208}\mathrm{Pb}$ UPCs at $\sqrt{s_{\rm NN}} = 5.5\,\mathrm{TeV}$. Cuts on kinematical variables of photons corresponding to the ATLAS experiment were imposed: $p_T > 2.5 \, {\rm GeV}, \, -2.4 < y < 2.4$. The resonant tetraquark contributions are: tensor $2^{++}$ $1s$ at $6.593\,\rm{GeV}$  (solid orange),
tensor $2^{++}$ $2s$ at $6.928\,\rm{GeV}$  (solid green),
scalar $0^{++}$ $1s$ doublet $6.435+6.542\,\rm{GeV}$  (dashed orange), and scalar $0^{++}$ $2s$ doublet $6.849+6.940\,\rm{GeV}$ (dashed green).
}
\label{fig:dsig_LBL}
\end{figure}
Figure~\ref{fig:dsig_LBL} displays the diphoton invariant-mass distribution on a logarithmic scale. The QED box continuum (solid blue) dominates across the full mass range while all of the resonant tetraquark contributions are small, at the level of a few $\times 10^{-2}\,\rm{nb}$, and similar in magnitude. The scalar $0^{++}$ doublets, after summing the two color eigenstates in each, yield results roughly twice as large as their tensor counterparts, as each individual scalar eigenstate has a magnitude comparable to the corresponding single tensor state. This reflects the constructive interference of the color configurations and the larger two-photon widths of the scalar states, notably $\Gamma_{\gamma\gamma}=0.688\,\rm{keV}$ for the $6940\,\rm{MeV}$, which compensate for the absence of the $(2J+1)=5$ spin factor.

\section{Conclusions}
\label{sec:conclusions}
In the present paper we have calculated $\gamma \gamma$ decay widths of fully charmed tetraquarks in a NRQCD approach. We have evaluated the relevant LDMEs for $0^{++}$ and $2^{++}$ states using quark model wave functions from Ref.\cite{Lu:2020cns}. We differ from earlier similar calculations in accounting for all available radial excitations $1s,2s,3s$, and in the assignment to the observed $X(6600),X(6900),X(7100)$ resonances, which we identify with $2^{++}(1s,2s,3s)$, respectively. The obtained LDMEs may be used in the future for other processes that can be addressed within NRQCD factorization.

As examples, in the present paper we have considered the production of fully heavy tetraquarks in the $J/\psi J/\psi$ and $\gamma \gamma$ final-state channels in ultraperipheral heavy ion collisions at the LHC.

There were claims in the literature that the fully charmed tetraquarks may be responsible for enhanced production of diphotons in $\gamma \gamma \to \gamma \gamma$ scattering in UPCs. Our calculation, based on realistic four-body wave functions for the $cc\bar{c}\bar{c}$ system obtained within the extended relativized quark model~\cite{Lu:2020cns} and combined with the NRQCD factorization framework, demonstrates that the resonant tetraquark contributions are more than two orders of magnitude smaller than the standard QED box continuum. Such a small signal-to-background ratio renders the $\gamma\gamma$ channel unpromising for tetraquark searches with current UPC luminosities.

In contrast, we have found that the tetraquark signal in the $J/\psi J/\psi$ channel exceeds the $\gamma\gamma \to J/\psi J/\psi$ continuum in the resonance region, making this channel substantially more promising. Whether a measurement in the $\mu^+ \mu^- \mu^+ \mu^-$ final state is feasible requires dedicated detector-level studies incorporating acceptance, trigger, and background considerations. As a benchmark, under the rather conservative assumption of a branching fraction $B_\psi \sim 3 \times 10^{-3}$, the resonant cross section for $X(6900) \to J/\psi J/\psi$ is of order $10\,\rm{nb}$.  With the typical integrated Pb--Pb UPC luminosity of a few $\rm{nb}^{-1}$ taken up to now, only a handful of signal events in the four--muon channel would be expected. A significant increase in the UPC data sample at the High-Luminosity LHC would therefore be highly valuable to fully exploit the discovery potential of this channel.

\section*{Acknowledgments}
This work was supported by the Polish National Science Center Grant No. UMO-2023/49/B/ST2/03665.
We are indebted to Qi-Fang L\"u, Dian-Yong Chen, and Yu-Bing Dong,  for making the results of Ref.\cite{Lu:2020cns} available to us.

\bibliography{references_tetraquark.bib}

\begin{thebibliography}{45}%
\makeatletter
\providecommand \@ifxundefined [1]{%
 \@ifx{#1\undefined}
}%
\providecommand \@ifnum [1]{%
 \ifnum #1\expandafter \@firstoftwo
 \else \expandafter \@secondoftwo
 \fi
}%
\providecommand \@ifx [1]{%
 \ifx #1\expandafter \@firstoftwo
 \else \expandafter \@secondoftwo
 \fi
}%
\providecommand \natexlab [1]{#1}%
\providecommand \enquote  [1]{``#1''}%
\providecommand \bibnamefont  [1]{#1}%
\providecommand \bibfnamefont [1]{#1}%
\providecommand \citenamefont [1]{#1}%
\providecommand \href@noop [0]{\@secondoftwo}%
\providecommand \href [0]{\begingroup \@sanitize@url \@href}%
\providecommand \@href[1]{\@@startlink{#1}\@@href}%
\providecommand \@@href[1]{\endgroup#1\@@endlink}%
\providecommand \@sanitize@url [0]{\catcode `\\12\catcode `\$12\catcode `\&12\catcode `\#12\catcode `\^12\catcode `\_12\catcode `\%12\relax}%
\providecommand \@@startlink[1]{}%
\providecommand \@@endlink[0]{}%
\providecommand \url  [0]{\begingroup\@sanitize@url \@url }%
\providecommand \@url [1]{\endgroup\@href {#1}{\urlprefix }}%
\providecommand \urlprefix  [0]{URL }%
\providecommand \Eprint [0]{\href }%
\providecommand \doibase [0]{http://dx.doi.org/}%
\providecommand \selectlanguage [0]{\@gobble}%
\providecommand \bibinfo  [0]{\@secondoftwo}%
\providecommand \bibfield  [0]{\@secondoftwo}%
\providecommand \translation [1]{[#1]}%
\providecommand \BibitemOpen [0]{}%
\providecommand \bibitemStop [0]{}%
\providecommand \bibitemNoStop [0]{.\EOS\space}%
\providecommand \EOS [0]{\spacefactor3000\relax}%
\providecommand \BibitemShut  [1]{\csname bibitem#1\endcsname}%
\let\auto@bib@innerbib\@empty
\bibitem [{\citenamefont {Aaij}\ \emph {et~al.}(2020)\citenamefont {Aaij} \emph {et~al.}}]{LHCb:2020bwg}%
  \BibitemOpen
  \bibfield  {author} {\bibinfo {author} {\bibfnamefont {Roel}\ \bibnamefont {Aaij}} \emph {et~al.} (\bibinfo {collaboration} {LHCb}),\ }\bibfield  {title} {\enquote {\bibinfo {title} {{Observation of structure in the $J /\psi$ -pair mass spectrum}},}\ }\href {\doibase 10.1016/j.scib.2020.08.032} {\bibfield  {journal} {\bibinfo  {journal} {Sci. Bull.}\ }\textbf {\bibinfo {volume} {65}},\ \bibinfo {pages} {1983--1993} (\bibinfo {year} {2020})},\ \Eprint {http://arxiv.org/abs/2006.16957} {arXiv:2006.16957 [hep-ex]} \BibitemShut {NoStop}%
\bibitem [{\citenamefont {Aad}\ \emph {et~al.}(2023)\citenamefont {Aad} \emph {et~al.}}]{ATLAS:2023bft}%
  \BibitemOpen
  \bibfield  {author} {\bibinfo {author} {\bibfnamefont {Georges}\ \bibnamefont {Aad}} \emph {et~al.} (\bibinfo {collaboration} {ATLAS}),\ }\bibfield  {title} {\enquote {\bibinfo {title} {{Observation of an Excess of Dicharmonium Events in the Four-Muon Final State with the ATLAS Detector}},}\ }\href {\doibase 10.1103/PhysRevLett.131.151902} {\bibfield  {journal} {\bibinfo  {journal} {Phys. Rev. Lett.}\ }\textbf {\bibinfo {volume} {131}},\ \bibinfo {pages} {151902} (\bibinfo {year} {2023})},\ \Eprint {http://arxiv.org/abs/2304.08962} {arXiv:2304.08962 [hep-ex]} \BibitemShut {NoStop}%
\bibitem [{\citenamefont {Aad}\ \emph {et~al.}(2025)\citenamefont {Aad} \emph {et~al.}}]{ATLAS:2025nsd}%
  \BibitemOpen
  \bibfield  {author} {\bibinfo {author} {\bibfnamefont {Georges}\ \bibnamefont {Aad}} \emph {et~al.} (\bibinfo {collaboration} {ATLAS}),\ }\bibfield  {title} {\enquote {\bibinfo {title} {{Observation of structures in the $J/\psi+\psi(2S)$ mass spectrum with the ATLAS detector}},}\ }\href@noop {} {\  (\bibinfo {year} {2025})},\ \Eprint {http://arxiv.org/abs/2509.13101} {arXiv:2509.13101 [hep-ex]} \BibitemShut {NoStop}%
\bibitem [{\citenamefont {Hayrapetyan}\ \emph {et~al.}(2024)\citenamefont {Hayrapetyan} \emph {et~al.}}]{CMS:2023owd}%
  \BibitemOpen
  \bibfield  {author} {\bibinfo {author} {\bibfnamefont {Aram}\ \bibnamefont {Hayrapetyan}} \emph {et~al.} (\bibinfo {collaboration} {CMS}),\ }\bibfield  {title} {\enquote {\bibinfo {title} {{New Structures in the J/{\ensuremath{\psi}}J/{\ensuremath{\psi}} Mass Spectrum in Proton-Proton Collisions at s=13{\,}{\,}TeV}},}\ }\href {\doibase 10.1103/PhysRevLett.132.111901} {\bibfield  {journal} {\bibinfo  {journal} {Phys. Rev. Lett.}\ }\textbf {\bibinfo {volume} {132}},\ \bibinfo {pages} {111901} (\bibinfo {year} {2024})},\ \Eprint {http://arxiv.org/abs/2306.07164} {arXiv:2306.07164 [hep-ex]} \BibitemShut {NoStop}%
\bibitem [{\citenamefont {Hayrapetyan}\ \emph {et~al.}(2025)\citenamefont {Hayrapetyan} \emph {et~al.}}]{CMS:2025fpt}%
  \BibitemOpen
  \bibfield  {author} {\bibinfo {author} {\bibfnamefont {Aram}\ \bibnamefont {Hayrapetyan}} \emph {et~al.} (\bibinfo {collaboration} {CMS}),\ }\bibfield  {title} {\enquote {\bibinfo {title} {{Determination of the spin and parity of all-charm tetraquarks}},}\ }\href {\doibase 10.1038/s41586-025-09711-7} {\bibfield  {journal} {\bibinfo  {journal} {Nature}\ }\textbf {\bibinfo {volume} {648}},\ \bibinfo {pages} {58--63} (\bibinfo {year} {2025})},\ \Eprint {http://arxiv.org/abs/2506.07944} {arXiv:2506.07944 [hep-ex]} \BibitemShut {NoStop}%
\bibitem [{\citenamefont {Hayrapetyan}\ \emph {et~al.}(2026)\citenamefont {Hayrapetyan} \emph {et~al.}}]{CMS:2026tiu}%
  \BibitemOpen
  \bibfield  {author} {\bibinfo {author} {\bibfnamefont {Aram}\ \bibnamefont {Hayrapetyan}} \emph {et~al.} (\bibinfo {collaboration} {CMS}),\ }\bibfield  {title} {\enquote {\bibinfo {title} {{Observation of a family of all-charm tetraquarks}},}\ }\href@noop {} {\  (\bibinfo {year} {2026})},\ \Eprint {http://arxiv.org/abs/2602.02252} {arXiv:2602.02252 [hep-ex]} \BibitemShut {NoStop}%
\bibitem [{\citenamefont {Berezhnoy}\ \emph {et~al.}(2011)\citenamefont {Berezhnoy}, \citenamefont {Likhoded}, \citenamefont {Luchinsky},\ and\ \citenamefont {Novoselov}}]{Berezhnoy:2011xy}%
  \BibitemOpen
  \bibfield  {author} {\bibinfo {author} {\bibfnamefont {A.~V.}\ \bibnamefont {Berezhnoy}}, \bibinfo {author} {\bibfnamefont {A.~K.}\ \bibnamefont {Likhoded}}, \bibinfo {author} {\bibfnamefont {A.~V.}\ \bibnamefont {Luchinsky}}, \ and\ \bibinfo {author} {\bibfnamefont {A.~A.}\ \bibnamefont {Novoselov}},\ }\bibfield  {title} {\enquote {\bibinfo {title} {{Double J/psi-meson Production at LHC and 4c-tetraquark state}},}\ }\href {\doibase 10.1103/PhysRevD.84.094023} {\bibfield  {journal} {\bibinfo  {journal} {Phys. Rev. D}\ }\textbf {\bibinfo {volume} {84}},\ \bibinfo {pages} {094023} (\bibinfo {year} {2011})},\ \Eprint {http://arxiv.org/abs/1101.5881} {arXiv:1101.5881 [hep-ph]} \BibitemShut {NoStop}%
\bibitem [{\citenamefont {Maciu{\l}a}\ \emph {et~al.}(2021)\citenamefont {Maciu{\l}a}, \citenamefont {Sch{\"a}fer},\ and\ \citenamefont {Szczurek}}]{Maciula:2020wri}%
  \BibitemOpen
  \bibfield  {author} {\bibinfo {author} {\bibfnamefont {Rafa{\l}}\ \bibnamefont {Maciu{\l}a}}, \bibinfo {author} {\bibfnamefont {Wolfgang}\ \bibnamefont {Sch{\"a}fer}}, \ and\ \bibinfo {author} {\bibfnamefont {Antoni}\ \bibnamefont {Szczurek}},\ }\bibfield  {title} {\enquote {\bibinfo {title} {{On the mechanism of $T_{4c}$(6900) tetraquark production}},}\ }\href {\doibase 10.1016/j.physletb.2020.136010} {\bibfield  {journal} {\bibinfo  {journal} {Phys. Lett. B}\ }\textbf {\bibinfo {volume} {812}},\ \bibinfo {pages} {136010} (\bibinfo {year} {2021})},\ \Eprint {http://arxiv.org/abs/2009.02100} {arXiv:2009.02100 [hep-ph]} \BibitemShut {NoStop}%
\bibitem [{\citenamefont {Feng}\ \emph {et~al.}(2022)\citenamefont {Feng}, \citenamefont {Huang}, \citenamefont {Jia}, \citenamefont {Sang}, \citenamefont {Xiong},\ and\ \citenamefont {Zhang}}]{Feng:2020riv}%
  \BibitemOpen
  \bibfield  {author} {\bibinfo {author} {\bibfnamefont {Feng}\ \bibnamefont {Feng}}, \bibinfo {author} {\bibfnamefont {Yingsheng}\ \bibnamefont {Huang}}, \bibinfo {author} {\bibfnamefont {Yu}~\bibnamefont {Jia}}, \bibinfo {author} {\bibfnamefont {Wen-Long}\ \bibnamefont {Sang}}, \bibinfo {author} {\bibfnamefont {Xiaonu}\ \bibnamefont {Xiong}}, \ and\ \bibinfo {author} {\bibfnamefont {Jia-Yue}\ \bibnamefont {Zhang}},\ }\bibfield  {title} {\enquote {\bibinfo {title} {{Fragmentation production of fully-charmed tetraquarks at the LHC}},}\ }\href {\doibase 10.1103/PhysRevD.106.114029} {\bibfield  {journal} {\bibinfo  {journal} {Phys. Rev. D}\ }\textbf {\bibinfo {volume} {106}},\ \bibinfo {pages} {114029} (\bibinfo {year} {2022})},\ \Eprint {http://arxiv.org/abs/2009.08450} {arXiv:2009.08450 [hep-ph]} \BibitemShut {NoStop}%
\bibitem [{\citenamefont {Zhu}(2021)}]{Zhu:2020xni}%
  \BibitemOpen
  \bibfield  {author} {\bibinfo {author} {\bibfnamefont {Ruilin}\ \bibnamefont {Zhu}},\ }\bibfield  {title} {\enquote {\bibinfo {title} {{Fully-heavy tetraquark spectra and production at hadron colliders}},}\ }\href {\doibase 10.1016/j.nuclphysb.2021.115393} {\bibfield  {journal} {\bibinfo  {journal} {Nucl. Phys. B}\ }\textbf {\bibinfo {volume} {966}},\ \bibinfo {pages} {115393} (\bibinfo {year} {2021})},\ \Eprint {http://arxiv.org/abs/2010.09082} {arXiv:2010.09082 [hep-ph]} \BibitemShut {NoStop}%
\bibitem [{\citenamefont {Feng}\ \emph {et~al.}(2023)\citenamefont {Feng}, \citenamefont {Huang}, \citenamefont {Jia}, \citenamefont {Sang}, \citenamefont {Yang},\ and\ \citenamefont {Zhang}}]{Feng:2023agq}%
  \BibitemOpen
  \bibfield  {author} {\bibinfo {author} {\bibfnamefont {Feng}\ \bibnamefont {Feng}}, \bibinfo {author} {\bibfnamefont {Yingsheng}\ \bibnamefont {Huang}}, \bibinfo {author} {\bibfnamefont {Yu}~\bibnamefont {Jia}}, \bibinfo {author} {\bibfnamefont {Wen-Long}\ \bibnamefont {Sang}}, \bibinfo {author} {\bibfnamefont {De-Shan}\ \bibnamefont {Yang}}, \ and\ \bibinfo {author} {\bibfnamefont {Jia-Yue}\ \bibnamefont {Zhang}},\ }\bibfield  {title} {\enquote {\bibinfo {title} {{Inclusive production of fully charmed tetraquarks at the LHC}},}\ }\href {\doibase 10.1103/PhysRevD.108.L051501} {\bibfield  {journal} {\bibinfo  {journal} {Phys. Rev. D}\ }\textbf {\bibinfo {volume} {108}},\ \bibinfo {pages} {L051501} (\bibinfo {year} {2023})},\ \Eprint {http://arxiv.org/abs/2304.11142} {arXiv:2304.11142 [hep-ph]} \BibitemShut {NoStop}%
\bibitem [{\citenamefont {Celiberto}\ \emph {et~al.}(2024)\citenamefont {Celiberto}, \citenamefont {Gatto},\ and\ \citenamefont {Papa}}]{Celiberto:2024mab}%
  \BibitemOpen
  \bibfield  {author} {\bibinfo {author} {\bibfnamefont {Francesco~Giovanni}\ \bibnamefont {Celiberto}}, \bibinfo {author} {\bibfnamefont {Gabriele}\ \bibnamefont {Gatto}}, \ and\ \bibinfo {author} {\bibfnamefont {Alessandro}\ \bibnamefont {Papa}},\ }\bibfield  {title} {\enquote {\bibinfo {title} {{Fully charmed tetraquarks from LHC to FCC: natural stability from fragmentation}},}\ }\href {\doibase 10.1140/epjc/s10052-024-13345-w} {\bibfield  {journal} {\bibinfo  {journal} {Eur. Phys. J. C}\ }\textbf {\bibinfo {volume} {84}},\ \bibinfo {pages} {1071} (\bibinfo {year} {2024})},\ \Eprint {http://arxiv.org/abs/2405.14773} {arXiv:2405.14773 [hep-ph]} \BibitemShut {NoStop}%
\bibitem [{\citenamefont {Belov}\ \emph {et~al.}(2025)\citenamefont {Belov}, \citenamefont {Giachino},\ and\ \citenamefont {Santopinto}}]{Belov:2024qyi}%
  \BibitemOpen
  \bibfield  {author} {\bibinfo {author} {\bibfnamefont {Ilia}\ \bibnamefont {Belov}}, \bibinfo {author} {\bibfnamefont {Alessandro}\ \bibnamefont {Giachino}}, \ and\ \bibinfo {author} {\bibfnamefont {Elena}\ \bibnamefont {Santopinto}},\ }\bibfield  {title} {\enquote {\bibinfo {title} {{Fully charmed tetraquark production at the LHC experiments}},}\ }\href {\doibase 10.1007/JHEP01(2025)093} {\bibfield  {journal} {\bibinfo  {journal} {JHEP}\ }\textbf {\bibinfo {volume} {01}},\ \bibinfo {pages} {093} (\bibinfo {year} {2025})},\ \Eprint {http://arxiv.org/abs/2409.12070} {arXiv:2409.12070 [hep-ph]} \BibitemShut {NoStop}%
\bibitem [{\citenamefont {Celiberto}(2025)}]{Celiberto:2025ziy}%
  \BibitemOpen
  \bibfield  {author} {\bibinfo {author} {\bibfnamefont {Francesco~Giovanni}\ \bibnamefont {Celiberto}},\ }\bibfield  {title} {\enquote {\bibinfo {title} {{Fragmentation of fully heavy tetraquarks: The TQ4Q1.1 functions as a case study}},}\ }\href {\doibase 10.1103/375n-fw5h} {\bibfield  {journal} {\bibinfo  {journal} {Phys. Rev. D}\ }\textbf {\bibinfo {volume} {112}},\ \bibinfo {pages} {074041} (\bibinfo {year} {2025})},\ \Eprint {http://arxiv.org/abs/2507.09744} {arXiv:2507.09744 [hep-ph]} \BibitemShut {NoStop}%
\bibitem [{\citenamefont {Wang}\ and\ \citenamefont {Zhu}(2025)}]{Wang:2025hex}%
  \BibitemOpen
  \bibfield  {author} {\bibinfo {author} {\bibfnamefont {Yefan}\ \bibnamefont {Wang}}\ and\ \bibinfo {author} {\bibfnamefont {Ruilin}\ \bibnamefont {Zhu}},\ }\bibfield  {title} {\enquote {\bibinfo {title} {{Fully charm tetraquark production at hadronic collisions with gluon radiation effects}},}\ }\href@noop {} {\  (\bibinfo {year} {2025})},\ \Eprint {http://arxiv.org/abs/2510.02085} {arXiv:2510.02085 [hep-ph]} \BibitemShut {NoStop}%
\bibitem [{\citenamefont {Celiberto}\ \emph {et~al.}(2026)\citenamefont {Celiberto}, \citenamefont {Giannini}, \citenamefont {Gon{\c{c}}alves},\ and\ \citenamefont {Lima}}]{Celiberto:2025vra}%
  \BibitemOpen
  \bibfield  {author} {\bibinfo {author} {\bibfnamefont {Francesco~G.}\ \bibnamefont {Celiberto}}, \bibinfo {author} {\bibfnamefont {Andr{\'e}~V.}\ \bibnamefont {Giannini}}, \bibinfo {author} {\bibfnamefont {Victor~P.}\ \bibnamefont {Gon{\c{c}}alves}}, \ and\ \bibinfo {author} {\bibfnamefont {Yuri~N.}\ \bibnamefont {Lima}},\ }\bibfield  {title} {\enquote {\bibinfo {title} {{Fully charmed tetraquark production in forward rapidity pp collisions at LHC and FCC energies}},}\ }\href {\doibase 10.1103/tq47-w7jn} {\bibfield  {journal} {\bibinfo  {journal} {Phys. Rev. D}\ }\textbf {\bibinfo {volume} {113}},\ \bibinfo {pages} {054014} (\bibinfo {year} {2026})},\ \Eprint {http://arxiv.org/abs/2511.18984} {arXiv:2511.18984 [hep-ph]} \BibitemShut {NoStop}%
\bibitem [{\citenamefont {Baranov}\ \emph {et~al.}(2013)\citenamefont {Baranov}, \citenamefont {Cisek}, \citenamefont {K{\l}usek-Gawenda}, \citenamefont {Sch{\"a}fer},\ and\ \citenamefont {Szczurek}}]{Baranov:2012vu}%
  \BibitemOpen
  \bibfield  {author} {\bibinfo {author} {\bibfnamefont {Sergey}\ \bibnamefont {Baranov}}, \bibinfo {author} {\bibfnamefont {Anna}\ \bibnamefont {Cisek}}, \bibinfo {author} {\bibfnamefont {Mariola}\ \bibnamefont {K{\l}usek-Gawenda}}, \bibinfo {author} {\bibfnamefont {Wolfgang}\ \bibnamefont {Sch{\"a}fer}}, \ and\ \bibinfo {author} {\bibfnamefont {Antoni}\ \bibnamefont {Szczurek}},\ }\bibfield  {title} {\enquote {\bibinfo {title} {{The $\gamma \gamma \to J/\psi J/\psi$ reaction and the $J/\psi J/\psi$ pair production in exclusive ultraperipheral ultrarelativistic heavy ion collisions}},}\ }\href {\doibase 10.1140/epjc/s10052-013-2335-8} {\bibfield  {journal} {\bibinfo  {journal} {Eur. Phys. J. C}\ }\textbf {\bibinfo {volume} {73}},\ \bibinfo {pages} {2335} (\bibinfo {year} {2013})},\ \Eprint {http://arxiv.org/abs/1208.5917} {arXiv:1208.5917 [hep-ph]} \BibitemShut {NoStop}%
\bibitem [{\citenamefont {Yang}\ \emph {et~al.}(2025)\citenamefont {Yang}, \citenamefont {Chen},\ and\ \citenamefont {Long}}]{Yang:2025vcs}%
  \BibitemOpen
  \bibfield  {author} {\bibinfo {author} {\bibfnamefont {Hao}\ \bibnamefont {Yang}}, \bibinfo {author} {\bibfnamefont {Zi-Qiang}\ \bibnamefont {Chen}}, \ and\ \bibinfo {author} {\bibfnamefont {Bingwei}\ \bibnamefont {Long}},\ }\bibfield  {title} {\enquote {\bibinfo {title} {{Charmonium pair production in ultraperipheral collision}},}\ }\href {\doibase 10.1140/epjc/s10052-025-14542-x} {\bibfield  {journal} {\bibinfo  {journal} {Eur. Phys. J. C}\ }\textbf {\bibinfo {volume} {85}},\ \bibinfo {pages} {802} (\bibinfo {year} {2025})},\ \Eprint {http://arxiv.org/abs/2504.14850} {arXiv:2504.14850 [hep-ph]} \BibitemShut {NoStop}%
\bibitem [{\citenamefont {Gon{\c{c}}alves}\ and\ \citenamefont {Moreira}(2021)}]{Goncalves:2021ytq}%
  \BibitemOpen
  \bibfield  {author} {\bibinfo {author} {\bibfnamefont {Victor~P.}\ \bibnamefont {Gon{\c{c}}alves}}\ and\ \bibinfo {author} {\bibfnamefont {Bruno~D.}\ \bibnamefont {Moreira}},\ }\bibfield  {title} {\enquote {\bibinfo {title} {{Fully - heavy tetraquark production by $\gamma\gamma$ interactions in hadronic collisions at the LHC}},}\ }\href {\doibase 10.1016/j.physletb.2021.136249} {\bibfield  {journal} {\bibinfo  {journal} {Phys. Lett. B}\ }\textbf {\bibinfo {volume} {816}},\ \bibinfo {pages} {136249} (\bibinfo {year} {2021})},\ \Eprint {http://arxiv.org/abs/2101.03798} {arXiv:2101.03798 [hep-ph]} \BibitemShut {NoStop}%
\bibitem [{\citenamefont {Esposito}\ \emph {et~al.}(2021)\citenamefont {Esposito}, \citenamefont {Manzari}, \citenamefont {Pilloni},\ and\ \citenamefont {Polosa}}]{Esposito:2021ptx}%
  \BibitemOpen
  \bibfield  {author} {\bibinfo {author} {\bibfnamefont {Angelo}\ \bibnamefont {Esposito}}, \bibinfo {author} {\bibfnamefont {Claudio~Andrea}\ \bibnamefont {Manzari}}, \bibinfo {author} {\bibfnamefont {Alessandro}\ \bibnamefont {Pilloni}}, \ and\ \bibinfo {author} {\bibfnamefont {Antonio~Davide}\ \bibnamefont {Polosa}},\ }\bibfield  {title} {\enquote {\bibinfo {title} {{Hunting for tetraquarks in ultraperipheral heavy ion collisions}},}\ }\href {\doibase 10.1103/PhysRevD.104.114029} {\bibfield  {journal} {\bibinfo  {journal} {Phys. Rev. D}\ }\textbf {\bibinfo {volume} {104}},\ \bibinfo {pages} {114029} (\bibinfo {year} {2021})},\ \Eprint {http://arxiv.org/abs/2109.10359} {arXiv:2109.10359 [hep-ph]} \BibitemShut {NoStop}%
\bibitem [{\citenamefont {Biloshytskyi}\ \emph {et~al.}(2022)\citenamefont {Biloshytskyi}, \citenamefont {Pascalutsa}, \citenamefont {Harland-Lang}, \citenamefont {Malaescu}, \citenamefont {Schmieden},\ and\ \citenamefont {Schott}}]{Biloshytskyi:2022dmo}%
  \BibitemOpen
  \bibfield  {author} {\bibinfo {author} {\bibfnamefont {Volodymyr}\ \bibnamefont {Biloshytskyi}}, \bibinfo {author} {\bibfnamefont {Vladimir}\ \bibnamefont {Pascalutsa}}, \bibinfo {author} {\bibfnamefont {Lucian}\ \bibnamefont {Harland-Lang}}, \bibinfo {author} {\bibfnamefont {Bogdan}\ \bibnamefont {Malaescu}}, \bibinfo {author} {\bibfnamefont {Kristof}\ \bibnamefont {Schmieden}}, \ and\ \bibinfo {author} {\bibfnamefont {Matthias}\ \bibnamefont {Schott}},\ }\bibfield  {title} {\enquote {\bibinfo {title} {{Two-photon decay of X(6900) from light-by-light scattering at the LHC}},}\ }\href {\doibase 10.1103/PhysRevD.106.L111902} {\bibfield  {journal} {\bibinfo  {journal} {Phys. Rev. D}\ }\textbf {\bibinfo {volume} {106}},\ \bibinfo {pages} {L111902} (\bibinfo {year} {2022})},\ \Eprint {http://arxiv.org/abs/2207.13623} {arXiv:2207.13623 [hep-ph]} \BibitemShut {NoStop}%
\bibitem [{\citenamefont {d'Enterria}\ and\ \citenamefont {Kang}(2025)}]{dEnterria:2025ecx}%
  \BibitemOpen
  \bibfield  {author} {\bibinfo {author} {\bibfnamefont {David}\ \bibnamefont {d'Enterria}}\ and\ \bibinfo {author} {\bibfnamefont {Karen}\ \bibnamefont {Kang}},\ }\bibfield  {title} {\enquote {\bibinfo {title} {{Exclusive photon-fusion production of even-spin resonances and exotic QED atoms in high-energy hadron collisions}},}\ }\href {\doibase 10.1103/rnxl-v6gd} {\bibfield  {journal} {\bibinfo  {journal} {Phys. Rev. D}\ }\textbf {\bibinfo {volume} {112}},\ \bibinfo {pages} {116022} (\bibinfo {year} {2025})},\ \Eprint {http://arxiv.org/abs/2503.10952} {arXiv:2503.10952 [hep-ph]} \BibitemShut {NoStop}%
\bibitem [{\citenamefont {Sang}\ \emph {et~al.}(2024)\citenamefont {Sang}, \citenamefont {Wang}, \citenamefont {Zhang},\ and\ \citenamefont {Feng}}]{Sang:2023ncm}%
  \BibitemOpen
  \bibfield  {author} {\bibinfo {author} {\bibfnamefont {Wen-Long}\ \bibnamefont {Sang}}, \bibinfo {author} {\bibfnamefont {Tao}\ \bibnamefont {Wang}}, \bibinfo {author} {\bibfnamefont {Yu-Dong}\ \bibnamefont {Zhang}}, \ and\ \bibinfo {author} {\bibfnamefont {Feng}\ \bibnamefont {Feng}},\ }\bibfield  {title} {\enquote {\bibinfo {title} {{Electromagnetic and hadronic decay of fully heavy tetraquarks}},}\ }\href {\doibase 10.1103/PhysRevD.109.056016} {\bibfield  {journal} {\bibinfo  {journal} {Phys. Rev. D}\ }\textbf {\bibinfo {volume} {109}},\ \bibinfo {pages} {056016} (\bibinfo {year} {2024})},\ \Eprint {http://arxiv.org/abs/2307.16150} {arXiv:2307.16150 [hep-ph]} \BibitemShut {NoStop}%
\bibitem [{\citenamefont {Kalamidas}\ and\ \citenamefont {Vanderhaeghen}(2025)}]{Kalamidas:2025gen}%
  \BibitemOpen
  \bibfield  {author} {\bibinfo {author} {\bibfnamefont {Panagiotis}\ \bibnamefont {Kalamidas}}\ and\ \bibinfo {author} {\bibfnamefont {Marc}\ \bibnamefont {Vanderhaeghen}},\ }\bibfield  {title} {\enquote {\bibinfo {title} {{All-charm tetraquark and its contribution to two-photon processes}},}\ }\href {\doibase 10.1103/PhysRevD.111.094033} {\bibfield  {journal} {\bibinfo  {journal} {Phys. Rev. D}\ }\textbf {\bibinfo {volume} {111}},\ \bibinfo {pages} {094033} (\bibinfo {year} {2025})},\ \Eprint {http://arxiv.org/abs/2501.06034} {arXiv:2501.06034 [hep-ph]} \BibitemShut {NoStop}%
\bibitem [{\citenamefont {Becchi}\ \emph {et~al.}(2020)\citenamefont {Becchi}, \citenamefont {Ferretti}, \citenamefont {Giachino}, \citenamefont {Maiani},\ and\ \citenamefont {Santopinto}}]{Becchi:2020uvq}%
  \BibitemOpen
  \bibfield  {author} {\bibinfo {author} {\bibfnamefont {C.}~\bibnamefont {Becchi}}, \bibinfo {author} {\bibfnamefont {J.}~\bibnamefont {Ferretti}}, \bibinfo {author} {\bibfnamefont {A.}~\bibnamefont {Giachino}}, \bibinfo {author} {\bibfnamefont {L.}~\bibnamefont {Maiani}}, \ and\ \bibinfo {author} {\bibfnamefont {E.}~\bibnamefont {Santopinto}},\ }\bibfield  {title} {\enquote {\bibinfo {title} {{A study of $c c\bar{c}\bar{c}$ tetraquark decays in 4 muons and in $D^{(*)} \bar{D}^{(*)}$ at LHC}},}\ }\href {\doibase 10.1016/j.physletb.2020.135952} {\bibfield  {journal} {\bibinfo  {journal} {Phys. Lett. B}\ }\textbf {\bibinfo {volume} {811}},\ \bibinfo {pages} {135952} (\bibinfo {year} {2020})},\ \Eprint {http://arxiv.org/abs/2006.14388} {arXiv:2006.14388 [hep-ph]} \BibitemShut {NoStop}%
\bibitem [{\citenamefont {Zhang}\ \emph {et~al.}(2024)\citenamefont {Zhang}, \citenamefont {Mo},\ and\ \citenamefont {Yan}}]{Zhang:2023ffe}%
  \BibitemOpen
  \bibfield  {author} {\bibinfo {author} {\bibfnamefont {Hong-Fei}\ \bibnamefont {Zhang}}, \bibinfo {author} {\bibfnamefont {Xue-Mei}\ \bibnamefont {Mo}}, \ and\ \bibinfo {author} {\bibfnamefont {Yu-Peng}\ \bibnamefont {Yan}},\ }\bibfield  {title} {\enquote {\bibinfo {title} {{Hadronic decay of exotic mesons consisting of four charm quarks}},}\ }\href {\doibase 10.1103/PhysRevD.110.096021} {\bibfield  {journal} {\bibinfo  {journal} {Phys. Rev. D}\ }\textbf {\bibinfo {volume} {110}},\ \bibinfo {pages} {096021} (\bibinfo {year} {2024})},\ \Eprint {http://arxiv.org/abs/2312.10850} {arXiv:2312.10850 [hep-ph]} \BibitemShut {NoStop}%
\bibitem [{\citenamefont {Poppe}(1986)}]{Poppe:1986dq}%
  \BibitemOpen
  \bibfield  {author} {\bibinfo {author} {\bibfnamefont {M.}~\bibnamefont {Poppe}},\ }\bibfield  {title} {\enquote {\bibinfo {title} {{Exclusive Hadron Production in Two Photon Reactions}},}\ }\href {\doibase 10.1142/S0217751X8600023X} {\bibfield  {journal} {\bibinfo  {journal} {Int. J. Mod. Phys. A}\ }\textbf {\bibinfo {volume} {1}},\ \bibinfo {pages} {545--668} (\bibinfo {year} {1986})}\BibitemShut {NoStop}%
\bibitem [{\citenamefont {Budnev}\ \emph {et~al.}(1975)\citenamefont {Budnev}, \citenamefont {Ginzburg}, \citenamefont {Meledin},\ and\ \citenamefont {Serbo}}]{Budnev:1975poe}%
  \BibitemOpen
  \bibfield  {author} {\bibinfo {author} {\bibfnamefont {V.~M.}\ \bibnamefont {Budnev}}, \bibinfo {author} {\bibfnamefont {I.~F.}\ \bibnamefont {Ginzburg}}, \bibinfo {author} {\bibfnamefont {G.~V.}\ \bibnamefont {Meledin}}, \ and\ \bibinfo {author} {\bibfnamefont {V.~G.}\ \bibnamefont {Serbo}},\ }\bibfield  {title} {\enquote {\bibinfo {title} {{The Two photon particle production mechanism. Physical problems. Applications. Equivalent photon approximation}},}\ }\href {\doibase 10.1016/0370-1573(75)90009-5} {\bibfield  {journal} {\bibinfo  {journal} {Phys. Rept.}\ }\textbf {\bibinfo {volume} {15}},\ \bibinfo {pages} {181--281} (\bibinfo {year} {1975})}\BibitemShut {NoStop}%
\bibitem [{\citenamefont {Bodwin}\ \emph {et~al.}(1995)\citenamefont {Bodwin}, \citenamefont {Braaten},\ and\ \citenamefont {Lepage}}]{Bodwin:1994jh}%
  \BibitemOpen
  \bibfield  {author} {\bibinfo {author} {\bibfnamefont {Geoffrey~T.}\ \bibnamefont {Bodwin}}, \bibinfo {author} {\bibfnamefont {Eric}\ \bibnamefont {Braaten}}, \ and\ \bibinfo {author} {\bibfnamefont {G.~Peter}\ \bibnamefont {Lepage}},\ }\bibfield  {title} {\enquote {\bibinfo {title} {{Rigorous QCD analysis of inclusive annihilation and production of heavy quarkonium}},}\ }\href {\doibase 10.1103/PhysRevD.55.5853} {\bibfield  {journal} {\bibinfo  {journal} {Phys. Rev. D}\ }\textbf {\bibinfo {volume} {51}},\ \bibinfo {pages} {1125--1171} (\bibinfo {year} {1995})},\ \bibinfo {note} {[Erratum: Phys.Rev.D 55, 5853 (1997)]},\ \Eprint {http://arxiv.org/abs/hep-ph/9407339} {arXiv:hep-ph/9407339} \BibitemShut {NoStop}%
\bibitem [{\citenamefont {Shtabovenko}\ \emph {et~al.}(2025)\citenamefont {Shtabovenko}, \citenamefont {Mertig},\ and\ \citenamefont {Orellana}}]{Shtabovenko:2023idz}%
  \BibitemOpen
  \bibfield  {author} {\bibinfo {author} {\bibfnamefont {Vladyslav}\ \bibnamefont {Shtabovenko}}, \bibinfo {author} {\bibfnamefont {Rolf}\ \bibnamefont {Mertig}}, \ and\ \bibinfo {author} {\bibfnamefont {Frederik}\ \bibnamefont {Orellana}},\ }\bibfield  {title} {\enquote {\bibinfo {title} {{FeynCalc 10: Do multiloop integrals dream of computer codes?}}}\ }\href {\doibase 10.1016/j.cpc.2024.109357} {\bibfield  {journal} {\bibinfo  {journal} {Comput. Phys. Commun.}\ }\textbf {\bibinfo {volume} {306}},\ \bibinfo {pages} {109357} (\bibinfo {year} {2025})},\ \Eprint {http://arxiv.org/abs/2312.14089} {arXiv:2312.14089 [hep-ph]} \BibitemShut {NoStop}%
\bibitem [{\citenamefont {L{\"u}}\ \emph {et~al.}(2020)\citenamefont {L{\"u}}, \citenamefont {Chen},\ and\ \citenamefont {Dong}}]{Lu:2020cns}%
  \BibitemOpen
  \bibfield  {author} {\bibinfo {author} {\bibfnamefont {Qi-Fang}\ \bibnamefont {L{\"u}}}, \bibinfo {author} {\bibfnamefont {Dian-Yong}\ \bibnamefont {Chen}}, \ and\ \bibinfo {author} {\bibfnamefont {Yu-Bing}\ \bibnamefont {Dong}},\ }\bibfield  {title} {\enquote {\bibinfo {title} {{Masses of fully heavy tetraquarks $QQ {\bar{Q}} {\bar{Q}}$ in an extended relativized quark model}},}\ }\href {\doibase 10.1140/epjc/s10052-020-08454-1} {\bibfield  {journal} {\bibinfo  {journal} {Eur. Phys. J. C}\ }\textbf {\bibinfo {volume} {80}},\ \bibinfo {pages} {871} (\bibinfo {year} {2020})},\ \Eprint {http://arxiv.org/abs/2006.14445} {arXiv:2006.14445 [hep-ph]} \BibitemShut {NoStop}%
\bibitem [{\citenamefont {Hiyama}\ \emph {et~al.}(2003)\citenamefont {Hiyama}, \citenamefont {Kino},\ and\ \citenamefont {Kamimura}}]{Hiyama:2003cu}%
  \BibitemOpen
  \bibfield  {author} {\bibinfo {author} {\bibfnamefont {E.}~\bibnamefont {Hiyama}}, \bibinfo {author} {\bibfnamefont {Y.}~\bibnamefont {Kino}}, \ and\ \bibinfo {author} {\bibfnamefont {M.}~\bibnamefont {Kamimura}},\ }\bibfield  {title} {\enquote {\bibinfo {title} {{Gaussian expansion method for few-body systems}},}\ }\href {\doibase 10.1016/S0146-6410(03)90015-9} {\bibfield  {journal} {\bibinfo  {journal} {Prog. Part. Nucl. Phys.}\ }\textbf {\bibinfo {volume} {51}},\ \bibinfo {pages} {223--307} (\bibinfo {year} {2003})}\BibitemShut {NoStop}%
\bibitem [{\citenamefont {Chetyrkin}\ \emph {et~al.}(2000)\citenamefont {Chetyrkin}, \citenamefont {Kuhn},\ and\ \citenamefont {Steinhauser}}]{Chetyrkin:2000yt}%
  \BibitemOpen
  \bibfield  {author} {\bibinfo {author} {\bibfnamefont {K.~G.}\ \bibnamefont {Chetyrkin}}, \bibinfo {author} {\bibfnamefont {Johann~H.}\ \bibnamefont {Kuhn}}, \ and\ \bibinfo {author} {\bibfnamefont {M.}~\bibnamefont {Steinhauser}},\ }\bibfield  {title} {\enquote {\bibinfo {title} {{RunDec: A Mathematica package for running and decoupling of the strong coupling and quark masses}},}\ }\href {\doibase 10.1016/S0010-4655(00)00155-7} {\bibfield  {journal} {\bibinfo  {journal} {Comput. Phys. Commun.}\ }\textbf {\bibinfo {volume} {133}},\ \bibinfo {pages} {43--65} (\bibinfo {year} {2000})},\ \Eprint {http://arxiv.org/abs/hep-ph/0004189} {arXiv:hep-ph/0004189} \BibitemShut {NoStop}%
\bibitem [{\citenamefont {Badalian}\ \emph {et~al.}(1987)\citenamefont {Badalian}, \citenamefont {Ioffe},\ and\ \citenamefont {Smilga}}]{Badalian:1985es}%
  \BibitemOpen
  \bibfield  {author} {\bibinfo {author} {\bibfnamefont {A.~M.}\ \bibnamefont {Badalian}}, \bibinfo {author} {\bibfnamefont {B.~L.}\ \bibnamefont {Ioffe}}, \ and\ \bibinfo {author} {\bibfnamefont {Andrei~V.}\ \bibnamefont {Smilga}},\ }\bibfield  {title} {\enquote {\bibinfo {title} {{Four quark states in the heavy quark system}},}\ }\href {\doibase 10.1016/0550-3213(87)90248-3} {\bibfield  {journal} {\bibinfo  {journal} {Nucl. Phys. B}\ }\textbf {\bibinfo {volume} {281}},\ \bibinfo {pages} {85} (\bibinfo {year} {1987})}\BibitemShut {NoStop}%
\bibitem [{\citenamefont {Liu}\ \emph {et~al.}(2025)\citenamefont {Liu}, \citenamefont {Wang},\ and\ \citenamefont {Zhu}}]{Liu:2025mxv}%
  \BibitemOpen
  \bibfield  {author} {\bibinfo {author} {\bibfnamefont {Xinran}\ \bibnamefont {Liu}}, \bibinfo {author} {\bibfnamefont {Yefan}\ \bibnamefont {Wang}}, \ and\ \bibinfo {author} {\bibfnamefont {Ruilin}\ \bibnamefont {Zhu}},\ }\bibfield  {title} {\enquote {\bibinfo {title} {{Next-to-leading order QCD corrections to electromagnetic production and decay of fully charm tetraquarks}},}\ }\href@noop {} {\  (\bibinfo {year} {2025})},\ \Eprint {http://arxiv.org/abs/2512.22070} {arXiv:2512.22070 [hep-ph]} \BibitemShut {NoStop}%
\bibitem [{\citenamefont {Aad}\ \emph {et~al.}(2021)\citenamefont {Aad} \emph {et~al.}}]{ATLAS:2020hii}%
  \BibitemOpen
  \bibfield  {author} {\bibinfo {author} {\bibfnamefont {Georges}\ \bibnamefont {Aad}} \emph {et~al.} (\bibinfo {collaboration} {ATLAS}),\ }\bibfield  {title} {\enquote {\bibinfo {title} {{Measurement of light-by-light scattering and search for axion-like particles with 2.2 nb$^{-1}$ of Pb+Pb data with the ATLAS detector}},}\ }\href {\doibase 10.1007/JHEP03(2021)243} {\bibfield  {journal} {\bibinfo  {journal} {JHEP}\ }\textbf {\bibinfo {volume} {03}},\ \bibinfo {pages} {243} (\bibinfo {year} {2021})},\ \bibinfo {note} {[Erratum: JHEP 11, 050 (2021)]},\ \Eprint {http://arxiv.org/abs/2008.05355} {arXiv:2008.05355 [hep-ex]} \BibitemShut {NoStop}%
\bibitem [{\citenamefont {Debastiani}\ and\ \citenamefont {Navarra}(2019)}]{Debastiani:2017msn}%
  \BibitemOpen
  \bibfield  {author} {\bibinfo {author} {\bibfnamefont {V.~R.}\ \bibnamefont {Debastiani}}\ and\ \bibinfo {author} {\bibfnamefont {F.~S.}\ \bibnamefont {Navarra}},\ }\bibfield  {title} {\enquote {\bibinfo {title} {{A non-relativistic model for the $[cc][\bar{c}\bar{c}]$ tetraquark}},}\ }\href {\doibase 10.1088/1674-1137/43/1/013105} {\bibfield  {journal} {\bibinfo  {journal} {Chin. Phys. C}\ }\textbf {\bibinfo {volume} {43}},\ \bibinfo {pages} {013105} (\bibinfo {year} {2019})},\ \Eprint {http://arxiv.org/abs/1706.07553} {arXiv:1706.07553 [hep-ph]} \BibitemShut {NoStop}%
\bibitem [{\citenamefont {Baur}\ \emph {et~al.}(2002)\citenamefont {Baur}, \citenamefont {Hencken}, \citenamefont {Trautmann}, \citenamefont {Sadovsky},\ and\ \citenamefont {Kharlov}}]{Baur:2001jj}%
  \BibitemOpen
  \bibfield  {author} {\bibinfo {author} {\bibfnamefont {Gerhard}\ \bibnamefont {Baur}}, \bibinfo {author} {\bibfnamefont {Kai}\ \bibnamefont {Hencken}}, \bibinfo {author} {\bibfnamefont {Dirk}\ \bibnamefont {Trautmann}}, \bibinfo {author} {\bibfnamefont {Serguei}\ \bibnamefont {Sadovsky}}, \ and\ \bibinfo {author} {\bibfnamefont {Yuri}\ \bibnamefont {Kharlov}},\ }\bibfield  {title} {\enquote {\bibinfo {title} {{Coherent gamma gamma and gamma-A interactions in very peripheral collisions at relativistic ion colliders}},}\ }\href {\doibase 10.1016/S0370-1573(01)00101-6} {\bibfield  {journal} {\bibinfo  {journal} {Phys. Rept.}\ }\textbf {\bibinfo {volume} {364}},\ \bibinfo {pages} {359--450} (\bibinfo {year} {2002})},\ \Eprint {http://arxiv.org/abs/hep-ph/0112211} {arXiv:hep-ph/0112211} \BibitemShut {NoStop}%
\bibitem [{\citenamefont {Sch{\"a}fer}(2020)}]{Schafer:2020bnm}%
  \BibitemOpen
  \bibfield  {author} {\bibinfo {author} {\bibfnamefont {Wolfgang}\ \bibnamefont {Sch{\"a}fer}},\ }\bibfield  {title} {\enquote {\bibinfo {title} {{Photon induced processes: from ultraperipheral to semicentral heavy ion collisions}},}\ }\href {\doibase 10.1140/epja/s10050-020-00231-8} {\bibfield  {journal} {\bibinfo  {journal} {Eur. Phys. J. A}\ }\textbf {\bibinfo {volume} {56}},\ \bibinfo {pages} {231} (\bibinfo {year} {2020})}\BibitemShut {NoStop}%
\bibitem [{\citenamefont {Lu}\ and\ \citenamefont {Jiang}(2025)}]{Lu:2025lyu}%
  \BibitemOpen
  \bibfield  {author} {\bibinfo {author} {\bibfnamefont {Duo-Duo}\ \bibnamefont {Lu}}\ and\ \bibinfo {author} {\bibfnamefont {Shao-Zhou}\ \bibnamefont {Jiang}},\ }\bibfield  {title} {\enquote {\bibinfo {title} {{Investigating the internal structure of $X(6900)$ in the $2J/\psi$ decay channel}},}\ }\href@noop {} {\  (\bibinfo {year} {2025})},\ \Eprint {http://arxiv.org/abs/2512.18569} {arXiv:2512.18569 [hep-ph]} \BibitemShut {NoStop}%
\bibitem [{\citenamefont {liu}\ \emph {et~al.}(2024)\citenamefont {liu}, \citenamefont {Liu}, \citenamefont {Zhong},\ and\ \citenamefont {Zhao}}]{liu:2020eha}%
  \BibitemOpen
  \bibfield  {author} {\bibinfo {author} {\bibfnamefont {Ming-Sheng}\ \bibnamefont {liu}}, \bibinfo {author} {\bibfnamefont {Feng-Xiao}\ \bibnamefont {Liu}}, \bibinfo {author} {\bibfnamefont {Xian-Hui}\ \bibnamefont {Zhong}}, \ and\ \bibinfo {author} {\bibfnamefont {Qiang}\ \bibnamefont {Zhao}},\ }\bibfield  {title} {\enquote {\bibinfo {title} {{Fully heavy tetraquark states and their evidences in LHC observations}},}\ }\href {\doibase 10.1103/PhysRevD.109.076017} {\bibfield  {journal} {\bibinfo  {journal} {Phys. Rev. D}\ }\textbf {\bibinfo {volume} {109}},\ \bibinfo {pages} {076017} (\bibinfo {year} {2024})},\ \Eprint {http://arxiv.org/abs/2006.11952} {arXiv:2006.11952 [hep-ph]} \BibitemShut {NoStop}%
\bibitem [{\citenamefont {Aaboud}\ \emph {et~al.}(2017)\citenamefont {Aaboud} \emph {et~al.}}]{ATLAS:2017fur}%
  \BibitemOpen
  \bibfield  {author} {\bibinfo {author} {\bibfnamefont {Morad}\ \bibnamefont {Aaboud}} \emph {et~al.} (\bibinfo {collaboration} {ATLAS}),\ }\bibfield  {title} {\enquote {\bibinfo {title} {{Evidence for light-by-light scattering in heavy-ion collisions with the ATLAS detector at the LHC}},}\ }\href {\doibase 10.1038/nphys4208} {\bibfield  {journal} {\bibinfo  {journal} {Nature Phys.}\ }\textbf {\bibinfo {volume} {13}},\ \bibinfo {pages} {852--858} (\bibinfo {year} {2017})},\ \Eprint {http://arxiv.org/abs/1702.01625} {arXiv:1702.01625 [hep-ex]} \BibitemShut {NoStop}%
\bibitem [{\citenamefont {A~H}\ \emph {et~al.}(2024)\citenamefont {A~H}, \citenamefont {Chaubey}, \citenamefont {Fraaije}, \citenamefont {Hirschi},\ and\ \citenamefont {Shao}}]{AH:2023kor}%
  \BibitemOpen
  \bibfield  {author} {\bibinfo {author} {\bibfnamefont {Ajjath}\ \bibnamefont {A~H}}, \bibinfo {author} {\bibfnamefont {Ekta}\ \bibnamefont {Chaubey}}, \bibinfo {author} {\bibfnamefont {Mathijs}\ \bibnamefont {Fraaije}}, \bibinfo {author} {\bibfnamefont {Valentin}\ \bibnamefont {Hirschi}}, \ and\ \bibinfo {author} {\bibfnamefont {Hua-Sheng}\ \bibnamefont {Shao}},\ }\bibfield  {title} {\enquote {\bibinfo {title} {{Light-by-light scattering at next-to-leading order in QCD and QED}},}\ }\href {\doibase 10.1016/j.physletb.2024.138555} {\bibfield  {journal} {\bibinfo  {journal} {Phys. Lett. B}\ }\textbf {\bibinfo {volume} {851}},\ \bibinfo {pages} {138555} (\bibinfo {year} {2024})},\ \Eprint {http://arxiv.org/abs/2312.16956} {arXiv:2312.16956 [hep-ph]} \BibitemShut {NoStop}%
\bibitem [{\citenamefont {Bargiela}\ \emph {et~al.}(2026)\citenamefont {Bargiela}, \citenamefont {Chakraborty}, \citenamefont {Gambuti},\ and\ \citenamefont {Ozcelik}}]{Bargiela:2026tcn}%
  \BibitemOpen
  \bibfield  {author} {\bibinfo {author} {\bibfnamefont {Piotr}\ \bibnamefont {Bargiela}}, \bibinfo {author} {\bibfnamefont {Amlan}\ \bibnamefont {Chakraborty}}, \bibinfo {author} {\bibfnamefont {Giulio}\ \bibnamefont {Gambuti}}, \ and\ \bibinfo {author} {\bibfnamefont {Melih~A.}\ \bibnamefont {Ozcelik}},\ }\bibfield  {title} {\enquote {\bibinfo {title} {{Light-by-light scattering at three loops in massless QCD and QED: amplitudes and cross sections}},}\ }\href@noop {} {\  (\bibinfo {year} {2026})},\ \Eprint {http://arxiv.org/abs/2603.22423} {arXiv:2603.22423 [hep-ph]} \BibitemShut {NoStop}%
\bibitem [{\citenamefont {Bern}\ \emph {et~al.}(2001)\citenamefont {Bern}, \citenamefont {De~Freitas}, \citenamefont {Dixon}, \citenamefont {Ghinculov},\ and\ \citenamefont {Wong}}]{Bern:2001dg}%
  \BibitemOpen
  \bibfield  {author} {\bibinfo {author} {\bibfnamefont {Z.}~\bibnamefont {Bern}}, \bibinfo {author} {\bibfnamefont {A.}~\bibnamefont {De~Freitas}}, \bibinfo {author} {\bibfnamefont {Lance~J.}\ \bibnamefont {Dixon}}, \bibinfo {author} {\bibfnamefont {A.}~\bibnamefont {Ghinculov}}, \ and\ \bibinfo {author} {\bibfnamefont {H.~L.}\ \bibnamefont {Wong}},\ }\bibfield  {title} {\enquote {\bibinfo {title} {{QCD and QED corrections to light by light scattering}},}\ }\href {\doibase 10.1088/1126-6708/2001/11/031} {\bibfield  {journal} {\bibinfo  {journal} {JHEP}\ }\textbf {\bibinfo {volume} {11}},\ \bibinfo {pages} {031} (\bibinfo {year} {2001})},\ \Eprint {http://arxiv.org/abs/hep-ph/0109079} {arXiv:hep-ph/0109079} \BibitemShut {NoStop}%
\end{thebibliography}%

\end{document}